\documentclass[twocolumn]{aastex61}
\pdfoutput=1
\usepackage{tcolorbox}
\usepackage{savesym}
\usepackage{soul}

\usepackage{amsmath}
\savesymbol{tablenum}
\usepackage{verbatim}
\usepackage{siunitx}
\restoresymbol{SIX}{tablenum}
\definecolor{orcidlogocol}{HTML}{A6CE39}
\usepackage{natbib}
\usepackage{hyperref}
\usepackage{graphicx}
\usepackage{float}

\defcitealias{Bond2017}{B17}
\defcitealias{Shvartzvald2017}{S17}
\usepackage{xcolor}
\colorlet{RED}{red}

\begin{document}

	\title{OGLE-2016-BLG-1195L\lowercase{b}: A sub-Neptune Beyond the Snow Line of an M-dwarf Confirmed by Keck Adaptive Optics}

		\author[0000-0002-9881-4760]{Aikaterini Vandorou}
	\affil{Laboratory for Exoplanets and Stellar Astrophysics, NASA/Goddard Space Flight Center, Greenbelt, MD 20771, USA}
		\affil{Department of Astronomy, University of Maryland, College Park, MD 20742, USA}
			\affil{Center for Research and Exploration in Space Science and Technology, NASA/GSFC, Greenbelt, MD 20771}

		\author[0000-0003-4987-6591]{Lisa Dang}
		\affil{Department of Physics, McGill University,
		3600 University St, Montr\'{e}al, QC H3A 2T8, Canada}
	\affil{Universit\'e de Montr\'eal, Institut de Recherche sur les Exoplan\`etes, 1375 Ave.Th\'er\`ese-Lavoie-Roux, Montr\'eal, QC H2V 0B3}
	
			\author[0000-0001-8043-8413]{David P. Bennett}
		\affil{Laboratory for Exoplanets and Stellar Astrophysics, NASA/Goddard Space Flight Center, Greenbelt, MD 20771, USA}
		\affil{Department of Astronomy, University of Maryland, College Park, MD 20742, USA}
		\affil{Center for Research and Exploration in Space Science and Technology, NASA/GSFC, Greenbelt, MD 20771}
		
			\author[0000-0003-2302-9562]{Naoki Koshimoto}
				\affil{Department of Earth and Space Science, Graduate School of Science, Osaka University, Tokyonaka, Osaka, 560-0043, Japan}

		\author[0000-0002-5029-3257]{Sean K. Terry}
			\affil{Laboratory for Exoplanets and Stellar Astrophysics, NASA/Goddard Space Flight Center, Greenbelt, MD 20771, USA}
				\affil{Department of Astronomy, University of Maryland, College Park, MD 20742, USA}
		\affil{Department of Astronomy, University of California Berkeley, Berkeley, CA 94701, USA}

			\author{Andrzej Udalski}
\affil{Astronomical Observatory, University of Warsaw, Al. Ujazdowskie 4,00-478 Warszawa,Poland}

			\author[0000-0003-0014-3354]{Jean-Philippe Beaulieu}
		\affil{School of Natural Sciences, University of Tasmania,
		Private Bag 37 Hobart, Tasmania, 7001, Australia}
	\affil{Sorbonne Universit\'e, CNRS, Institut d'Astrophysique de Paris, IAP, F-75014, Paris, France}

		\author{Christophe Alard}
		\affil{Sorbonne Universit\'e, CNRS, Institut d'Astrophysique de Paris, IAP, F-75014, Paris, France}

		\author{Aparna Bhattacharya}
	\affil{Laboratory for Exoplanets and Stellar Astrophysics, NASA/Goddard Space Flight Center, Greenbelt, MD 20771, USA}
	\affil{Department of Astronomy, University of Maryland, College Park, MD 20742, USA}
	
			\author[0000-0001-5860-1157]{Joshua W. Blackman}
	\affil{School of Natural Sciences, University of Tasmania,
		Private Bag 37 Hobart, Tasmania, 7001, Australia}
	
	\author[0000-0002-8131-8891]{Ian A. Bond}
	\affil{Institute of Natural and Mathematical Sciences, Massey University, Auckland 0745,
		 New Zealand}
	
			\author{Tarik Bouchoutrouch-Ku}
	\affil{Department of Physics, McGill University,
		3600 University St, Montr\'{e}al, QC H3A 2T8, Canada}
	
	\author[0000-0003-0303-3855]{Andrew A. Cole}
	\affil{School of Natural Sciences, University of Tasmania,
		Private Bag 37 Hobart, Tasmania, 7001, Australia}
	
	\author[0000-0001-6129-5699]{Nicolas B. Cowan}
		\affil{Department of Physics, McGill University,
		3600 University St, Montr\'{e}al, QC H3A 2T8, Canada}
	\affil{Department of Earth \& Planetary Sciences, McGill University, 3450 rue University, Montr\'{e}al, QC, H3A 0E8, Canada
	}
	
	\author[0000-0002-7901-7213]{Jean-Baptiste Marquette}
	\affil{Laboratoire d'Astrophysique de Bordeaux, Univ. Bordeaux, CNRS, B18N, all\'ee Geoffroy Saint-Hilaire, 33615 Pessac, France}
		\affil{Sorbonne Universit\'e, CNRS, Institut d'Astrophysique de Paris, IAP, F-75014, Paris, France}

	\author[0000-0003-2388-4534]{Cl\'ement Ranc}
	\affil{Sorbonne Universit\'e, CNRS, Institut d'Astrophysique de Paris, IAP, F-75014, Paris, France}
\affil{Zentrum f{\"u}r Astronomie der Universit{\"a}t Heidelberg, Astronomisches Rechen-Institut, M{\"o}nchhofstr.\ 12-14, 69120 Heidelberg, Germany}
	
	\author[0000-0002-1530-4870]{Natalia E. Rektsini}
		\affil{School of Natural Sciences, University of Tasmania,
		Private Bag 37 Hobart, Tasmania, 7001, Australia}
			\affil{Sorbonne Universit\'e, CNRS, Institut d'Astrophysique de Paris, IAP, F-75014, Paris, France}

	\author{Sylvain Cetre}
    \affil{W. M. Keck Observatory, 65-1120 Mamalahoa Hwy, Kamuela, HI 96743}	
    
    \author{Jim Lyke}
    \affil{W. M. Keck Observatory, 65-1120 Mamalahoa Hwy, Kamuela, HI 96743}
    
    \author{Eduardo Marin}
	\affil{W. M. Keck Observatory, 65-1120 Mamalahoa Hwy, Kamuela, HI 96743}
	
	\author{Peter Wizinowich}
	\affil{W. M. Keck Observatory, 65-1120 Mamalahoa Hwy, Kamuela, HI 96743}
	%%ABSTRACT
	
	\begin{abstract}

We present the analysis of high resolution follow-up observations of OGLE-2016-BLG-1195 using Laser Guide Star Adaptive Optics with Keck, seven years after the event's peak. We resolve the lens, measuring its flux and the relative source-lens proper motion, thus finding the system to be a $M_{\rm p} = 10.08\pm 1.18\ M_{\rm \oplus}$ planet  orbiting an M-dwarf, $M_{\rm L} = 0.62\pm 0.05\ M_{\odot}$, beyond the snow line, with a projected separation of $r_\perp=2.24\pm 0.21$ AU at $D_{\rm L} = 7.45\pm 0.55$ kpc. Our results are consistent with the discovery paper, which reports values with 1-sigma uncertainties based on a single mass-distance constraint from finite source effects. However, both the discovery paper and our follow-up results disagree with the analysis of a different group that also present the planetary signal detection. The latter utilizes \textit{Spitzer} photometry to measure a parallax signal claiming the system is an Earth-mass planet orbiting an ultracool dwarf. Their parallax signal though is improbable since it suggests a lens star in the disk moving perpendicular to or counter to the Galactic disk rotation. Moreover, microlensing parallaxes can be impacted by systematic errors in the photometry. Therefore, we reanalyze the \textit{Spitzer} photometry using a Pixel Level Decorrelation (PLD) model to detrend detector systematics. We find that we can not confidently recover the same detrended light-curve that is likely dominated by systematic errors in the photometric data. The results of this paper act as a cautionary tale that a careful understanding of detector systematics and how they influence astrophysical constraints is crucial.

	\end{abstract}
	
	%%KEYWORDS
	\keywords{adaptive optics - planets and satellites, gravitational lensing, detection - proper motions, parallax}

	\section{Introduction} \label{sec:intro}

Gravitational microlensing is a unique method that can detect Earth to super-Jupiter mass exoplanets in wide orbits beyond the snow line \citep{Bennett1996, Mao1991, Gould1992}. The snow line refers to a region just beyond the habitable zone where water can only be sustained in the form of ice, and it is the predicted region where more massive planets form \citep{Lissauer1993, Ida2004, Kennedy2006}.  Statistical studies have shown that the frequency of planets beyond the snow line (especially ice and gas giants), is about 7 times greater than smaller separation systems \citep{Gould2010, Cumming2008}. 

Contrary to other planetary detection techniques, microlensing can detect planets without using any light from the host star. This allows for a range of stellar type hosts to be probed, e.g. Solar-type, M dwarfs and white dwarfs \citep{Blackman2021}. In addition, this method can investigate more distant planetary systems in our galaxy towards the Galactic Bulge, thus complementing the other detection techniques that typically probe the Solar neighborhood.

The advantages of microlensing, however, also bring certain caveats. Not directly detecting the light from the host star means that the physical parameters of the system, such as mass and distance, cannot be determined easily. Instead, the method accurately constrains the host-planet mass ratio, which may be considered a more fundamental property of planetary systems than the planetary mass \citep{Pascucci2018,IAU2022}. However, the planet formation process is thought to depend sensitively on the 
host star and planet masses \citep{Mulders2015}, so both masses and mass ratios are important.

The individual masses of a planetary microlensing event can be determined if both second-order effects on the light-curve are measured (e.g. \cite{Furusawa2013, Udalski2018}); finite source effects (features due to the source radius) and the microlensing parallax, $\pi_E$. Most planetary microlensing events exhibit finite source effects, however, observable parallax effects are uncommon. Using just the finite source effects, the physical parameters can be estimated using a Bayesian analysis with a Galactic model as a prior \citep{Beaulieu2006, Gaudi2008}{\bf }. However, this method typically yields poorly constrained parameters, since it only uses a single constraint. Parameters can be estimated with greater precision when additional constraints on the mass-distance relation are added.

A complementary mass-measurement method that has been successful, is high angular resolution follow-up observations from 8-10m class ground telescopes, or the Hubble Space Telescope. Follow-up observations can take place several years after the microlensing event has occurred, which means the source and lens can be resolved and their fluxes measured individually (e.g. \cite{Bhattacharya2018, Vandorou2020, Bennett2020, Terrry2021, Terry2022}). With this method it is also important to identify the lens star, since companions to either the lens and source, can also be mistaken for the lens system. The source star can be identified by comparing the candidate lens-source separations to the relative proper motion predicted by the light-curve model.

In this paper we present the Keck follow-up observations of OGLE-2016-BLG-1195 (henceforth, OB161195) used to resolve the source and lens. Section~\ref{OGLE} discusses the previous analysis of this event. Section~\ref{Keck_obs} discusses the Keck adaptive optics
observations and our measurements of the flux ratio and separation of the lens and source, and constrain the relative lens-source proper motion to constrain the relative source-lens proper motion, thus obtaining accurate physical properties of the planetary system.
Section~\ref{sec:newmodel} describes the image-constrained light-curve modeling used \citep{Bennett2024} to find light-curve models that fit both the light-curve and high angular resolution follow-up data. Section~\ref{planetary system parameters} presents the inferred properties of the OGLE-2016-BLG-1195L planetary system, and we present our conclusions in Section~\ref{sec:conclude}.

 \begin{deluxetable*}{ l l l l l}
	\setlength\tabcolsep{18pt} 
	\tablecaption{Best fit light-curve model parameters from \citetalias{Bond2017} and \citetalias{Shvartzvald2017}, and the new values found from remodeling using a 2L1S (two lens one source) model and a 2L2S (two lens two source) model. The 2L1S and 2L2S models use image constrained modeling from the 2023 Keck data. The \citetalias{Shvartzvald2017} model has eight degenerate solution, we present the best-fit wide solution. Parameter descriptions can be found in Section \ref{sec:newmodel}. \label{BondSHVparams} }
	\tablewidth{0pt}
	\tablehead{
		\colhead{Parameters}	& \colhead{Bond17}  & \colhead{Shvartzvald17} & \colhead{2L1S}& \colhead{2L2S}
	}  % end header.
	%\hline
	%\hline
	\startdata
	$t_{\rm E}$ (days)& 10.16$\pm$ 0.25 & 9.95 $\pm$ 0.11 & 10.09 $\pm$ 0.11 & 9.72 $\pm$ 0.28 \\   
	$t_0$ (HJD') & 7568.772 $\pm$ 0.002 &  7568.769 $\pm$ 0.001 & 7568.771 $\pm$ 0.001 & 7568.771 $\pm$ 0.001 \\
	$t_{\rm fix}$ (HJD')  & -- & -- & 7569.0& 7569.0 \\
	$u_0$ &  0.0514 $\pm$ 0.0014 & 0.0532 $\pm$ 0.0007& 0.0522 $\pm$ 0.0007 & 0.0543  $\pm$ 0.0018 \\
	s & 1.0698 $\pm$ 0.0078 & $1.0862^{+0.0080}_{-0.0075} $& 1.0788 $\pm$ 0.0461 & 1.0798 $\pm$ 0.0446  \\
	$\alpha$ (rad) & 0.966 $\pm$ 0.005 & 0.969 $\pm$ 0.002 & 2.174 $\pm$ 0.002 & 2.173 $\pm$ 0.002\\
	$t_*$ (days) & 0.0336 $\pm$ 0.0023&$0.0283^{+0.0034}_{-0.0038}$ & 0.0310 $\pm$ 0.0017 & 0.0311 $\pm$ 0.0016   \\
	$q/10^{-5}$ & 4.25 $\pm$ 0.67 & $5.68^{+0.80}_{-0.72}$ & 4.80 $\pm$ 0.49 & 4.95 $\pm$ 0.49 \\
	$\theta_{\rm E}$ (mas)& 0.261 $\pm$ 0.020 & $0.286^{+0.053}_{-0.038}$ & 0.289 $\pm$ 0.008 & 0.290 $\pm$ 0.013  \\
	$\pi_{\rm E,N}$ & -  &  -0.3016 $\pm$0.0074  & 0.0187 $\pm$ 0.0059  & 0.0188 $\pm$ 0.0057 \\
	$\pi_{\rm E,E}$ & -  &  -0.377 $\pm$ 0.032  & 0.0552 $\pm$ 0.0052  & 0.0536 $\pm$ 0.0052\\
	$I_{\rm s,0}$ & 17.82 $\pm$ 0.02&  17.84 $\pm$ 0.04&  17.95 $\pm$ 0.04  &  $17.90 \pm 0.04$ \\
	$(V-I)_{\rm s,0}$ & 0.71 $\pm$ 0.02 & 0.68 $\pm$ 0.03 & 0.66 $\pm$ 0.06 & $0.66 \pm 0.05$\\
	$I_{\rm s2,0}$ & --&  --&  --&  $18.07 \pm 0.25$ \\
	$(V-I)_{\rm s2,0}$ & -- & -- & -- & $ 0.40 \pm 0.47$\\
	$t_{0,2}$ (HJD') & -- & -- & -- & 7584.089 $\pm$ 11.073 \\
	$u_{0,2}$ & --&-- & --& 2.4520 $\pm$ 0.4844 \\
	f2rI & -- & -- & -- & 0.701$\pm$ 0.226  \\ 
	f2rV & -- & -- & -- & 0.884 $\pm$ 0.302  \\
	$\rm dt_{\rm E21}$ (days) & -- & -- & -- & $0.243 \pm 0.219$ \\
	$\rm d\alpha$ (rad) & -- & -- & -- & $0.00244 \pm 0.022$\\
	$D_{\rm S}$ (kpc) & -- & -- & 8.75 $\pm$ 0.62 & 8.74 $\pm$ 0.63 \\
	$\chi^2$ & 19580.4 & 10213 & 28180.23& 28165.64 \\
	d.o.f & 19587& -  & 28170& 28175 \\
	\enddata
\end{deluxetable*}

	\section{Microlensing Event OGLE-2016-BLG-1195}
	\label{OGLE}
	
   The microlensing event OB161195 was alerted by the Optical Gravitational Lens Experiment (OGLE, \cite{Udalski1994}, \cite{Udalski2003}) on June 27 2016. The event was also independently alerted by the Microlensing Observations in Astrophysics (MOA, \cite{Bond2001}) collaboration on June 28 2016, where a 2.5 hour photometric perturbation was discovered (the event has the alternative name of MOA-2016-BLG-350). Observations were taken in the I and V band with OGLE, and R and V band with MOA. OB161195 is located towards the Galactic Bulge at (R.A., Dec) = (17:55:23.50, --30:12:26.1) and Galactic coordinates $(l,\ b )=(-0.004, -2.475)$. 
   
   The initial data, modeling and analyses of the microlensing light-curve is presented in \cite{Bond2017} (henceforth, \citetalias{Bond2017}). The event exhibited a planetary anomaly with a mass-ratio, $q = 4.2 \pm 0.7 \times 10^{-5}$, which at the time of detection was the lowest mass-ratio found in microlensing. The Einstein ring radius ($\theta_E$) was derived from the microlensing parameters (Einstein radius crossing time, $t_E$, and the source radius crossing time, $t_*$) and the angular size measurement of the source star, $\theta_{*}$. The physical parameters of the lensing system were estimated by employing a Bayesian technique, as described in \citet{Bennett2008}. A Galactic model was used which comprised of a double-exponential disk \citep{Reid2002} and a bar model \citep{Han1995}. The Bayesian estimate of the lens mass was found to be $M_{\rm L} = 0.37^{+0.38}_{-0.21}\ M_{\odot}$, at a distance of $D_{\rm L} = 7.20^{+0.80}_{-1.02}$ kpc. These values imply a planet mass of $M_{\rm planet} = 5.10^{+5.5}_{-2.85}\ M_{\rm \oplus}$. The high uncertainties of these values are due to there only being a single mass-distance constraint, the $\theta_{\rm E}$ value, therefore the Bayesian result is highly dependent on the prior assumption that stars of all masses are equally likely to host a plant at the measured $q$ and $s$. Therefore, these estimates could be wrong if this assumption is wrong.
   
   From the light-curve model \citetalias{Bond2017} find the apparent color and magnitude of the source star to be $(V-I, I)_{ \rm s} = (2.113, 19.581) \pm (0.020, 0.008) $. Using extinction values $A_I=1.627 \pm 0.018$, $E(V-I) = 1.450 \pm 0.007$ and  $A_K = 0.256 \pm 0.017$ \citep{Nishiyama2006, Surot2019}, and the color relation $ V- K = 1.53$ \citep{Bessell1988}, we find a predicted source magnitude of $K_{\rm s, pred} = 17.34 \pm 0.04$ mag. This can be compared at a later stage to our measure Keck $K$-band value for the source.
   
   In parallel, the event was also observed in the paw-print of the Korean Microlensing Telescope Network (KMTNet, \cite{Kim2016}), which operates three 1.6m telescopes equipped with 4 square degree wide field imagers. Their data were taken in survey mode with the majority in \textit{I} band, and only a few in \textit{V}, and it was not influenced by the OGLE and MOA alert. Their data confirmed a planetary perturbation in the light-curve which was presented in \citet{Shvartzvald2017} (henceforth, \citetalias{Shvartzvald2017}). Their source star characteristics, extinction determination, mass-ratio and projected separations from the light-curve model are in agreement with \citetalias{Bond2017}. The light-curve parameters from both detection papers can be found in Table \ref{BondSHVparams}.

   Subsequent \textit{Spitzer} observations were also obtained with a cadence of one observation per day. A strong microlens parallax of $\pi_{\rm E} \sim$ 0.45 was retrieved by fitting the KMTNet and \textit{Spitzer} light-curves simultaneously, however, the low signal-to-noise \textit{Spitzer} data (presented in Figure 1 of \citetalias{Shvartzvald2017}) is seen to systematically differ from the model. In addition, \citetalias{Shvartzvald2017} report a lens-source relative proper motion of $\mu_{\rm hel, (N,E)} \equiv (\pm 4.0, -7.5)$ mas/yr measured from their parallax value, which suggests a counter-rotating lens in the disk. 
   
Combining this \textit{Spitzer} parallax with the light-curve model parameters from the KMTNet observations, \citetalias{Shvartzvald2017} found the system to be an Earth-mass planet, $M_{\rm planet} = 1.43^{+0.45}_{-0.32} M_{\rm \oplus} $, orbiting an ultra-cool dwarf, $M_{\rm L} = 0.078^{+0.016}_{-0.012} M_{\odot}$, at a distance of $D_{\rm L} = 3.19^{+0.42}_{-0.46}$ kpc. The large difference between the conclusions of \citetalias{Bond2017} and \citetalias{Shvartzvald2017} are entirely due to the \textit{Spitzer} data and their constraint on the parallax measurement. This measurement, and the reason for the discrepancy, is investigated in Section \ref{Spitzer}.

\section{Keck Follow-up of OGLE-2016-BLG-1195}
\label{Keck_obs}

We observed OB161195 with Keck II's laser guide star AO system \citep{Wizinowich2006} and the NIRC2 science instrument on August 3 2018. We used the wide camera with a plate scale of $\sim$39.69 mas/pixel. Images were taken using the $K_{\rm short}$ (henceforth, $K$) passband, where 15 good images were obtained. The exposure time of the images was 30 seconds with a dither of $\sim 2''$.  We reduced the images using the standard techniques described in \citet{Beaulieu2016} and \citet{Batista2014}. This procedure includes dark current and flat-field correction. After sky correcting the wide-frame images, they were stacked using SWARP \citep{Bertin2010} and the stellar fluxes were measured using aperture photometry with SExtractor \citep{Bertin1996}. We cross identify 71 stars in our stacked wide-frame with the Vista Variables in the Via Lactea (VVV) survey data \citep{Minniti2010} as described in \cite{Beaulieu2018}. This allowed us to calibrate our \textit{K} band Keck photometry. Thus, the total measured combined \textit{K} band magnitude of both source and lens is: 
\begin{equation*}
K_{\rm SL} =  16.38 \pm 0.04\ \rm mag.
\end{equation*}

 On August 12 2023, OB161195 was observed with Keck's OSIRIS instrument, and made use of the new Keck I real-time controller system as part of the KAPA upgrade \citep{Chin2022, wizinowich2022keck}. OSIRIS has a narrow camera with plate scale of 9.942 mas/pixel. 15 usable images were obtained in $\rm K_{prime}$ (hereafter, $\rm K_p$) band and reduced using the KAI (Keck AO Imaging) data reduction pipeline \citep{Lu_code_2022}. The FWHM (full-width-half-maximum) across each frame was $\sim 53$ mas with an average Strehl ratio of 0.368. The 14 frames were stacked using a jackknife method (\citet{Quenouille1949, Quenouille1956, Tukey1958}), where $\rm N-1$ frames are stacked, where N is the number of total frames (for this event N=15). Therefore, we produce 15 different stacks of 14 frames, since one frame is removed each time. See section \ref{mcmc+jkkf} for more details on the jackknife routine and getting error estimates using this method.

 The predicted separation between the source and lens is estimated to be $\sim$ 67 mas in 2023 ($\sim 7.12 $ years after peak magnification) with a predicted geocentric source-lens relative proper motion of $\sim \mu_{\rm Geo, rel}=9.38 \pm 0.75$ mas/yr \citep{Bond2017}. Since this separation is less than 1 FWHM, we expect the source and lens to be partially resolved, and thus found it necessary to use a point spread function (PSF) fitting photometry routine to measure both stars individually.

Additional epochs of data for this event were also collected in 2019 , 2020 and 2021. 
The quality of the 2019 images were very low with a FWHM of $> 100$ mas. At a predicted separation of $\sim$ 33 mas the lens would be undetectable and therefore, the data were not used.  The quality of the 2020 data was good, with a FWHM of $\sim$ 60 mas, and the lens was faintly detected. The quality of the 2021 data was also not very high with a FWHM of $\sim$ 75 mas. The lens was not visibly detected by eye, however, an MCMC revealed a tentative lens detection. The additional epochs of data are presented in Section \ref{sec:epochs} The 2018 wide frame images  were used for calibration only, whereas the 15 jackknife frames from 2023 (and additional data from 2020 and 2021) were used to resolve the source and lens. The detailed analysis of the narrow images are described in Section \ref{psf}.

\subsection{Target Identification}

\begin{figure*}[t!]
	\centering
	\includegraphics[width=14cm]{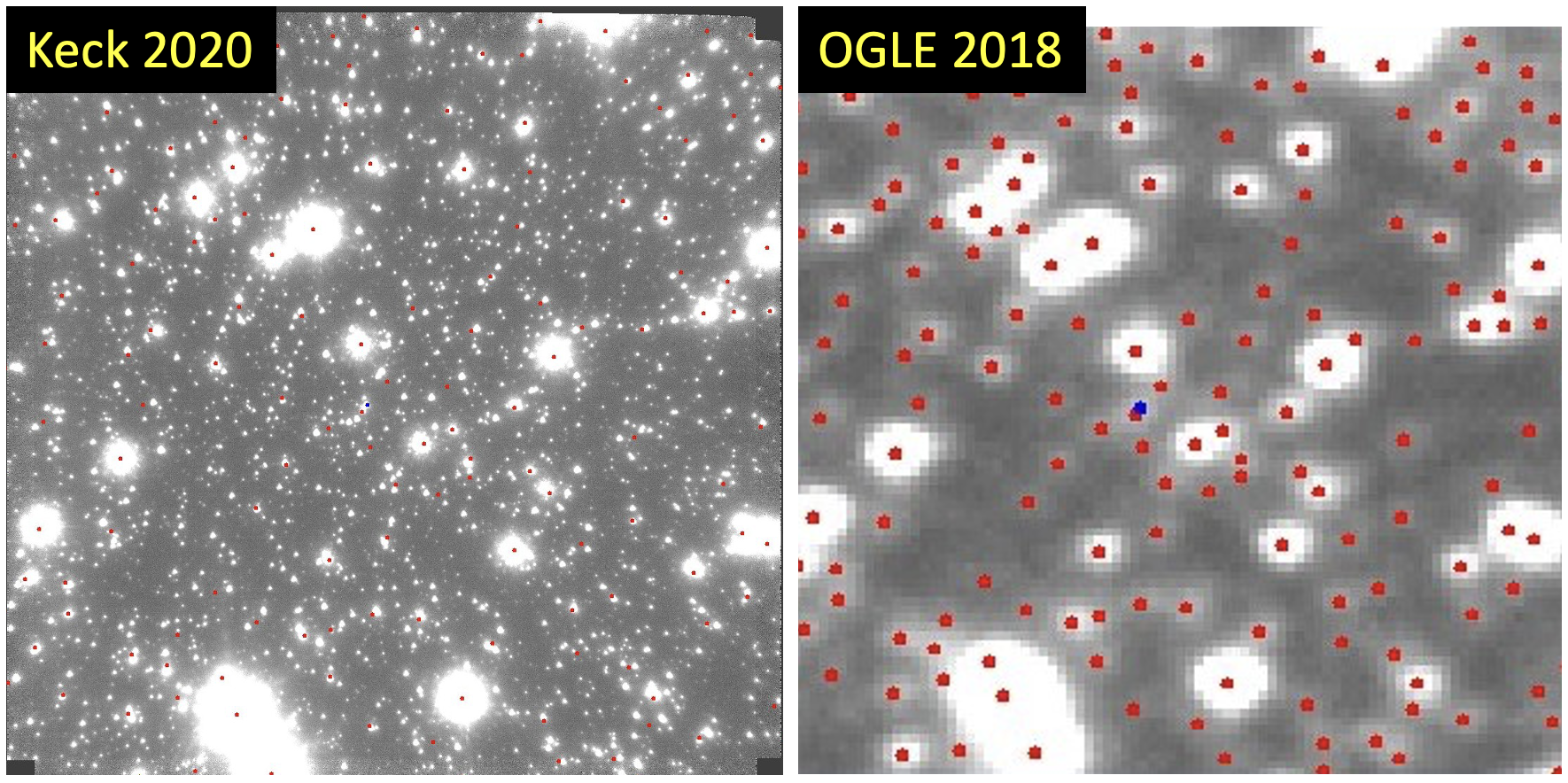}
	\caption{OGLE identification of OB161195. Red dots on the OGLE image are OGLE detected stars. The target is shown as a blue dot, which is the cetroid measured at the peak of magnification. In the OGLE reference image it is a blend of faint stars. The Keck image shows the OGLE stars tranformed to the Keck grid, in red, and the target in blue. The Keck OSIRIS frame is 9.942 mas/pixel, and the OGLE frame is 260 mas/pixel.}
	\label{fig:ogle_id}
\end{figure*}

\begin{figure*}[t!]
	\centering
	\includegraphics[width=14cm]{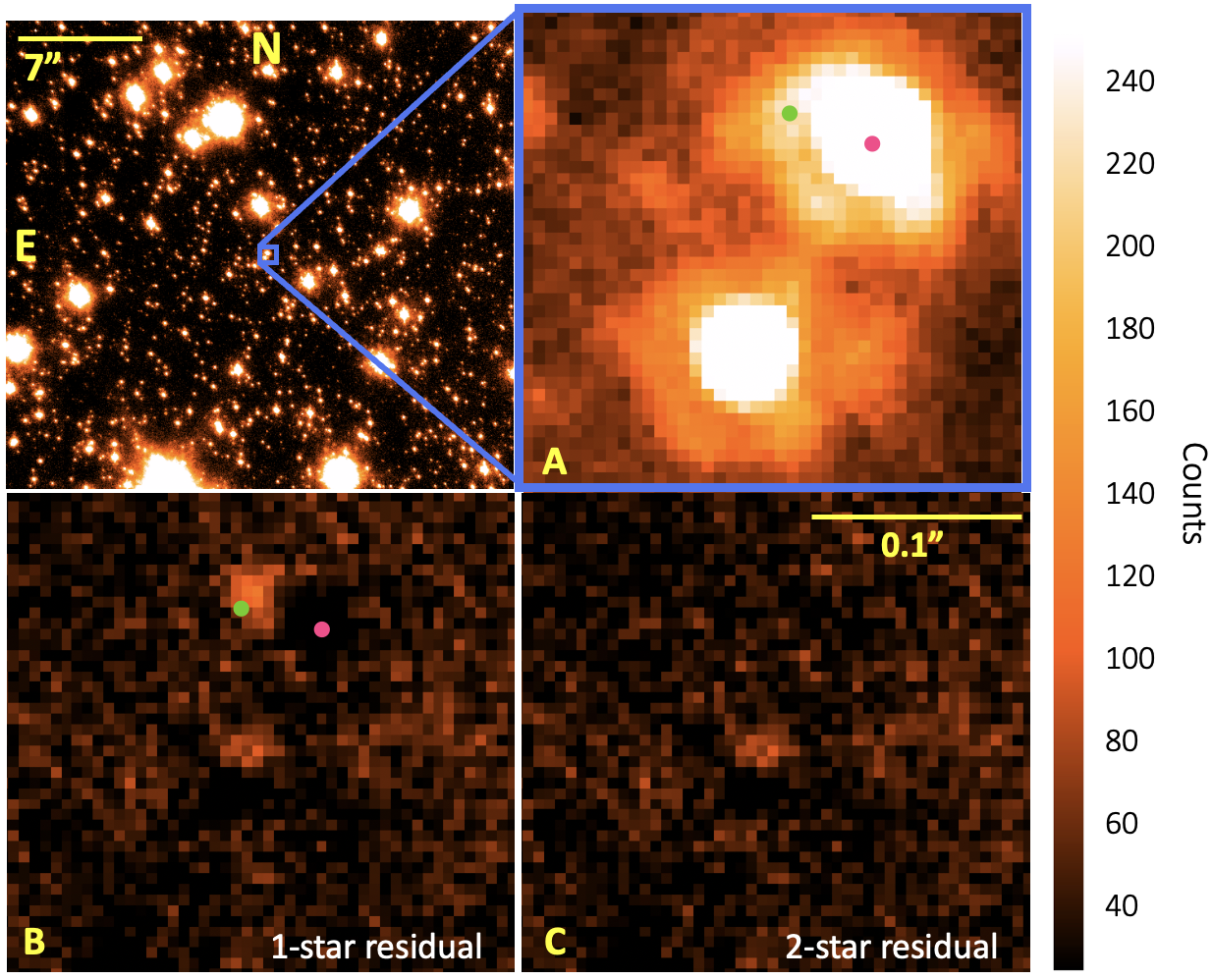}
	\caption{Keck OSIRIS K-band jackknife image from the 2023 observing run. The images have a pixel size of $\sim$9.942 mas/pixel. Panel A marks the target as the pink dot and the approximate position of the residual with a green dot. Panel B shows the residuals from a 1-star fit, and panel C shows the smooth residuals from a 2-star fit. All stars in the frame are fitted with the PSF model and subracted to produce the residual frame.}
	\label{fig:1195residualdaophot}
\end{figure*}

The initial analysis conducted with the 2020 data was with the target being misidentified as the star directly South East of the true target (see Figure \ref{fig:1195residualdaophot} panel A - the correct target is identified with a red dot).  This problem was first brought to the author's attention by \cite{Gould2023}. The misidentified target star did reveal a residual feature in the 2020 data which was thought to be the lens. This faint feature was again seen in the 2023 data, however, the proper motion was not consistent with it being the lens (see Appendix \ref{app:A} for more details.). 

The misidetification was also confounded by the fact that the false target had the predicted source brightness, while no other star in the vicinity did. However, given that a xallarap model was evaluated (but ultimately rejected) in the discovery paper by \cite{Bond2017}, we should have considered the possibility of a binary source, and therefore the posibility that the target would be ``brighter" than predicted.

To confirm which star is the true OB161195 target we conducted a reanalysis of the MOA data. This gave inconclusive results for the identification between the false and correct target, as different referece images favored different stars. Finally, the correct target was identified in the OGLE data using multiple reference images, and the result was conclusive (see Figure \ref{fig:ogle_id}).

\subsection{PSF Fitting on 2023 data}
\label{psf}

On each of the 15 jackknife frames we used a PSF fitting routine to detect the lens, which we expect to be partially resolved and relatively faint compared to the source. The analysis is done using a modified version of the DAOPHOT-II  package \citep{Stetson1987}. We call this \textit{DAOPHOT-MCMC}, as it runs Markov Chain Monte Carlo (MCMC) sampling on the target star. Additional details of this MCMC routine are outlined in \cite{Terrry2021}.

The first step in the \textit{DAOPHOT-MCMC} routine is to build an empirical PSF model using  single bright stars in the field of the 15 jackknife frames. The stars chosen are located toward the center of the field where image distortion is the least. Using the same reference stars, a different PSF model is built for each of the jackknife frames because the shape of the PSF varies from image to image. These variations contribute to the astrometric and photometric errors described further in Section \ref{mcmc+jkkf}. 

We tested three PSF models, each of which had a different combination of reference stars. The selected stars had to be bright, but not saturated, single and close to the area of the target star which is at the center of the frame. This narrowed our list down to 3-5 stars, therefore there is some overlap between the models. To determine which PSF model to use, we compared their ``chi" values, which is the normalised root-mean-square of the residuals that are left. PSF model 1 had a chi = 0.0543, model 2 a chi = 0.0533, and model 3 a chi = 0.0553. Therefore, we continued with model 2, which had the lowest value.

The PSF model is then fit to the target star, which is a blend of the source and lens. This produces corresponding residual frames for each of the 15 jackknife images, and can be seen in panel B of Figure \ref{fig:1195residualdaophot}, which we call the ``1-star" residual. In this residual frame, all stars have been fit with the PSF model. Our residual frames indicate a signal which we identify as the lens star (Figure \ref{fig:1195residualdaophot}, panel B). We then rerun the routine on the 15 jackknife frames using the 2-star fitting mode on our target star, and this produces a featureless residual as shown in panel C of Figure \ref{fig:1195residualdaophot}.

Once we have identified the lens star, we use the MCMC portion of the \textit{DAOPHOT-MCMC} to constrain the (x,y) pixel positions of the source and lens star, as well as their flux ratio. We do this initially using a 2-star fit which assumes 1 source and 1 lens. However, this scenario implies that the source star would be almost 1 magnitude brighter than the predicted source K band magnitude from the discovery papers \citetalias{Bond2017}  and \citetalias{Shvartzvald2017} of $K_{\rm s,pred}= 17.34 \pm 0.04$ mag. This implies that there is very likely to be a binary companion to the source star. Therefore, we also explore a scenario where a third star is contributing flux, repeating our MCMC run using a 3-star fit which assumes 2 sources and 1 lens.

The \textit{DAOPHOT-MCMC} fitting routine outputs the best-fit values from each jackknife frame for lens and source star position, and their flux ratio. From these parameters we can obtain the source-lens relative proper motion (direction and magnitude), and calibrated $K$-band magnitudes for each star (see Table \ref{table:2} for the 2-star scenario and Table \ref{table:3} for the 3-star).

\subsubsection*{2-star scenario}

In the 2-star scenario (1 lens and 1 source), we find the direction of relative heliocentric proper motion to be $(\mu_{\rm Hel,East}, \mu_{\rm Hel,North}) = (10.393, 3.592) \pm (0.351, 0.383)$ mas/yr, and the vector $\vec{\mu}_{\rm rel, Hel}=10.995 \pm 0.355$ mas/yr. Using the flux ratio $f_L/f_S = 0.044 \pm 0.003 $, and the total flux that we measure from the wide-frame Keck images ($K_{\rm SL}=16.38 \pm 0.04$), we can calculate the flux of the lens. We do this by using the following equations: 
	\begin{equation}
		\label{1K}
		K_{\rm L} - K_{\rm S} = -2.5\log_{10}(f_{\rm L}/f_{\rm S})
	\end{equation}
	\begin{equation}
		\label{2K}
		K_{\rm SL}= -2.5\log_{10}(10^{-0.4K_{\rm L}}+10^{-0.4K_{\rm S}}),
	\end{equation}
	where the notations `L' and `S' represent the lens and source, respectively, and $K_{\rm Keck}$ is the K band magnitude of the target. Solving equations \ref{1K} and \ref{2K} simultaneously, we find a new lens and source $K$-band magnitudes of:
	
\begin{gather*}
	K_{\rm L} = 19.801 \pm 0.101\ \rm mag\\
	K_{\rm S} = 16.443 \pm 0.075\ \rm mag.
\end{gather*}
	
We notice that the $K_{\rm S}$ value is almost 1 mag brighter than the predicted $K_{\rm S, pred}$ from \citetalias{Bond2017}'s light-curve model parameters - hence motivating us to investigate a companion to the source.

\subsubsection*{3-star scenario}

In the 3-star scenario we consider a lens and 2 source stars which we name S1 and S2. The separation between these two source stars is $6.318 \pm 1.737$ mas. This is less than 1 pixel width (9.942 mas), therefore, it is unclear which is the primary source in the microlensing model. Both S1 and S2 have a similar separation to the lens, however, since S1 has the better constrained X and Y pixel positions we proceed with these values. Therefore, we find the direction of relative heliocentric proper motion to be $(\mu_{\rm Hel,East}, \mu_{\rm Hel,North}) = (10.505, 3.676) \pm (0.437, 0.600)$ mas/yr, and the vector $\mu_{\rm rel, Hel}=11.131 \pm 0.458$ mas/yr.

Using the flux ratio $f_L/f_{S1+S2} = 0.044 \pm 0.003 $, and the total flux that we know from the wide-frame Keck images ($K_{\rm SL}=16.38 \pm 0.04$), we can calculate the flux of the lens and the flux of the individual sources following the method described above. Therefore we find:

	\begin{gather*}
	K_{\rm L} = 19.817 \pm 0.077\ \rm mag \\
	K_{\rm  S1} = 16.958 \pm 0.141\ \rm mag\\
	K_{\rm  S2} = 17.453 \pm 0.180\ \rm mag.\\
	\end{gather*}

The large error bars on the S1 and S2 magnitudes relative to the lens are due to the very small separation between them, and not being resolved. A result of the very small separation (less than 1 pixel width) is that flux values are exchanged during the MCMC run since the position of each source is also exchanged. The lens star position and flux, however, is unaffected by the unconstrained positions and fluxes of the two source stars.

\subsubsection{Errors with $\rm MCMC+Jackknife$ }
\label{mcmc+jkkf}

A major source of uncertainty in the Keck images comes from optical distortion effects in the atmosphere. This can affect the PSF of individual stars in each frame and across a single image. By stacking the 2023 narrow-frames using the jackknife method, and running the PSF fitting routine on each, we can capture the errors associated with PSF variations in the individual images. This routine must be included to account for the systematic errors in the PSF model shapes. The jackknife method involves taking $N$ images, and forming a series of stacked images that consist of $N-1$ images. Therefore, since we have 15 images total, we produce 15 stacks that consist of of $N-1=14$ frames. Finally, the best-fit values and errors from each MCMC run are  combined using the jackknife error calculation, where the standard error of parameter $x$ can be found using: 

\begin{equation}
	\label{jack}
	SE(x) = \sqrt{\frac{N-1}{N} \sum(x_i - \overline{x})^2},
\end{equation}
where N is the number of stacked images, $x_i$ is the parameter measured in each of the stacked images, and $\overline{x}$ is the mean of $x$ for all N stacked images. Although the jackknife method takes into account he variations in PSF shapes throughout the different frames, it does not necesserily account for photon noise since we only have 15 images. Therefore, the uncertainty estimate is not more precise that $1/\sqrt{N}$.

The errors from each MCMC can also be combined by taking their average. Since this combination method yields a different error value compared to combining with the jackknife equation, we add these errors in quadrature for the most conservative result. This process is applied to both the 2-star scenario and the 3-star scenario.

\begin{deluxetable}{ l  r  }
	\setlength\tabcolsep{12pt} 
	\tablecaption{ Separation, relative proper motion, and flux ratio from the 2023 Keck jackknife images. Values represent the MCMC best fit values and errors from \textit{DAOPHOT-MCMC} 2-star fit analysis (section \ref{psf}).  }  \label{table:2}
	%\tablewidth{0pt} 
	\tablehead{	 Parameter & Value
	}
	\startdata
	$\rm S-L\ Sep$ (mas) & 78.291  $\pm$ 2.525   \\ 
	\hline
	$\rm S-L\ \mu_{rel, Hel, East}$ (mas/yr) & 10.393  $\pm$ 0.351 \\
	$\rm S-L\ \mu_{rel, Hel, North}$ (mas/yr) & 3.592 $\pm$ 0.383\\
	\hline
	L flux / S flux & 0.0447  $\pm$ 0.0027 \\
	\hline
	$\chi^2$ & 343 \\
	dof & 431 \\
	\enddata
\end{deluxetable}	

\begin{deluxetable}{ l  r  }
	\setlength\tabcolsep{12pt} 
	\tablecaption{Separation, relative proper motion, and flux ratio from the 2023 Keck jackknife images. Values represent the MCMC best fit values and errors from \textit{DAOPHOT-MCMC} 3-star fit analysis (section \ref{psf}).  }	\label{table:3} 	
	%\tablewidth{0pt} 
		\tablehead{	 Parameter & Value
		}
	\startdata
	$\rm S1-S2\ Sep$ (mas) & 6.318 $\pm$ 1.737   \\ 
	 $\rm S1-L\ Sep$ (mas) & 78.926  $\pm$ 3.136   \\ 
	 $\rm S2-L\ Sep$ (mas) &78.831  $\pm$ 3.746  \\ 
	 \hline
	$\rm S1-L\ \mu_{rel, Hel, East}$ (mas/yr) & 10.505  $\pm$ 0.437 \\
	$\rm S1-L\ \mu_{rel, Hel, North}$ (mas/yr) & 3.676 $\pm$ 0.600\\
	\hline
	L flux / S1+S2 flux & 0.0440  $\pm$ 0.0028 \\
	\hline
	$\chi^2$ & 342 \\
	dof & 428 \\
	\enddata
\end{deluxetable}

\begin{figure*}[t!]
	\centering
	\includegraphics[width=14cm]{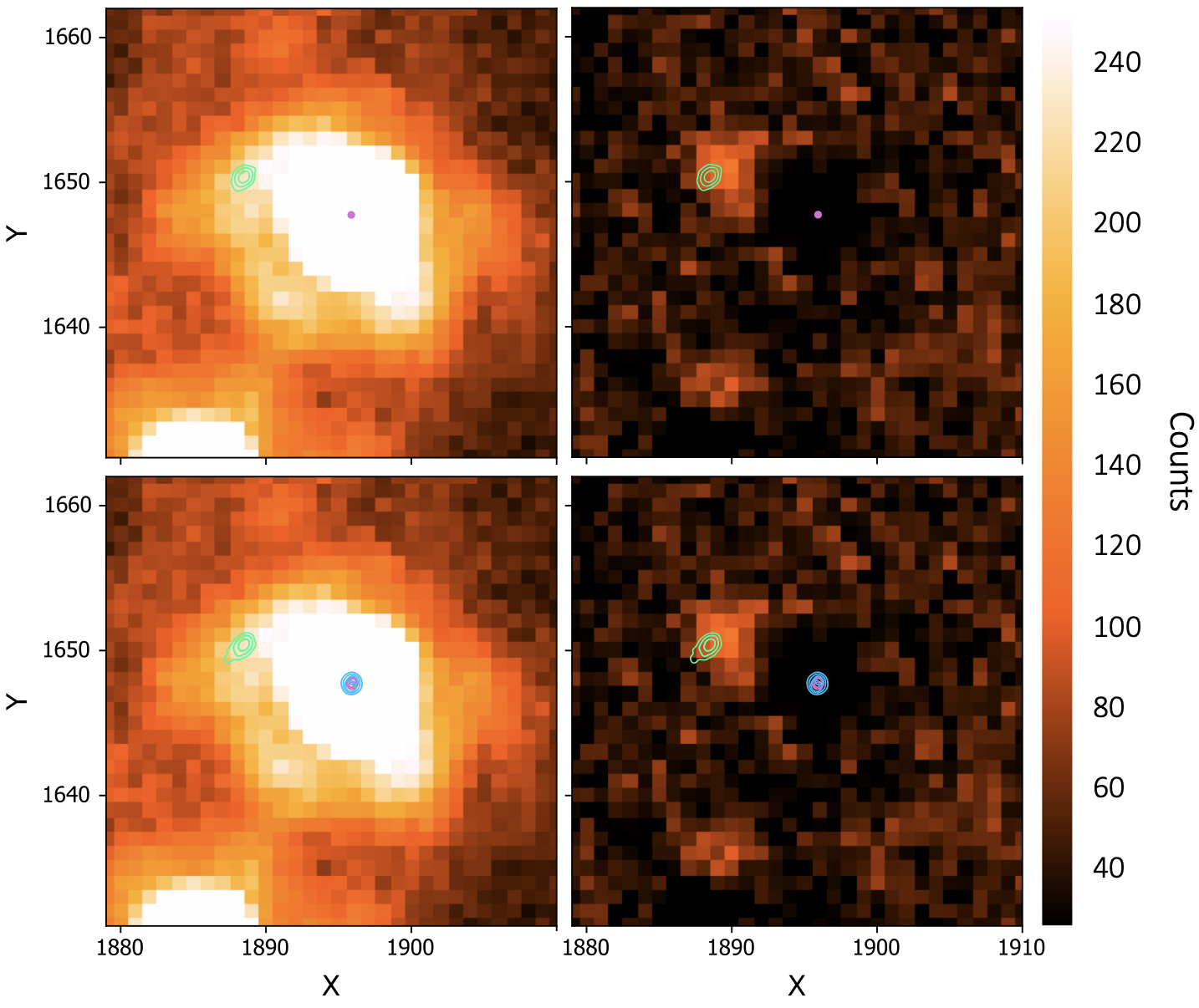}
	\caption{  Best fit MCMC contours ($68.3\%$, $99.5\%$, $99.7\%$) of stellar X and Y pixel positions. Top panels: lens (green) and 1 source star (pink). Bottom panels: lens (green) and 2 source stars (pink and blue). These contours are over-plotted on the 2023 jackknife K-band OSIRIS image (left), and the residuals of that image (right). Each frame is $\sim$ 0.35 $ \times$ 0.35  arcseconds.}
	\label{fig:contours}
\end{figure*}

\subsection{Additional Keck epochs}
\label{sec:epochs}

The Keck data from 2020 and 2021 were analysed using the same PSF fitting method described for the 2023 data. Using the \textit{DAOPHOT-MCMC} code we fit a 2-star PSF model. Since fitting the second source does not impact the relative proper motion or the lens flux, we do not find it necessary to fit a 3-star PSF model.   

The 2020 data was taken on August 8, 4.12 years after the peak of the microlensing event. The data consists of 11 images with the Keck I OSIRIS instrument, with a FWHM of $\sim 60$ mas across the frames and an average Strehl ratio of 0.343. We build a PSF model using 7 nearby stars in the frame, and then fit this model to the target star. We produce a 1-star residual where the source star is subtracted and see a very faint residual in the predicted position of the lens. We run an MCMC to determine the best fit source-lens separation and flux ratio. From the source-lens relative proper motion we measure in the 2023 data, we predict a separation of $\sim 45$ mas in the 2020 data, and use this as a constraint while running the MCMC. 

\begin{figure}[t!]
	\centering
	\includegraphics[width=8.5cm]{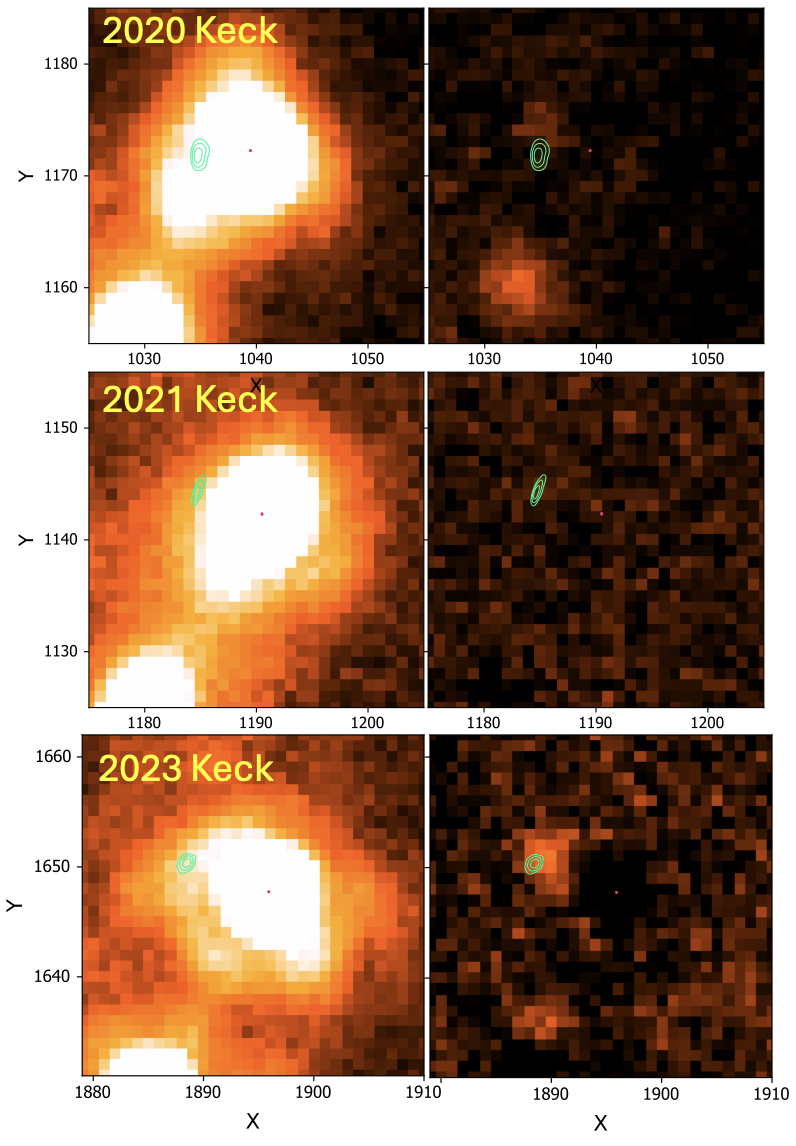}
	\caption{ Best fit MCMC contours of the source and lens pixel positions (X, Y) for 2020, 2021 and 2023. The left panels show the target, and the right panels show the 1-star residual with the source subtracted. The lens best fit positions are in green, and the source in red. }
	\label{fig:epoch_contours}
\end{figure}

A constrained MCMC was additionally run on the 2021 data. This data was also taken with the OSIRIS Keck I instrument, on August 14, 5.13 years after the even't peak. Across the 5 frames used we find a FWHM of $\sim 70$ mas and a Strehl ratio of 0.223. A predicted separation of $\sim 55$ mas is expected to be seen in the 2021 data, however, no visible detection of the lens is made upon inspecting the 1-star residual frame. A constrained MCMC is still run on this data using the predicted separation, with the results shown on Figure \ref{fig:epoch_contours}. 

It should be noted that the constraint used on the 2020 and 2021 data is just on the magnitude of the separation - not on the direction. In addition, we do not place any constraints on the source or lens flux.

\begin{deluxetable*}{ l  c  c  c }
	\setlength\tabcolsep{12pt} 
	\tablecaption{ Separation, relative proper motion, and flux ratio from the 2020 and  2021 Keck data compared to the 2023 data for the 2L1S model.  }  \label{table:epochs}
	%\tablewidth{0pt} 
	\tablehead{	 Parameter & 2020 data & 2021 data & 2023 data
	}
	\startdata
	Sep (mas) & 48.625 $\pm$ 4.022 & 44.661 $\pm$ 18.652 & 79.014  $\pm$ 2.807   \\ 
	$\rm \mu_{rel, Hel}$ (mas/yr) & 11.802 $\pm$ 0.482 & 10.864 $\pm$ 3.305 &10.393  $\pm$ 0.351 \\
	L flux / S flux & 0.039 $\pm$ 0.004 & 0.029 $\pm$ 0.007 &0.044  $\pm$ 0.003 \\
	Lens $K$ mag & 19.926 $\pm$ 0.108 & 20.259 $\pm$ 0.251 &  19.801 $\pm$ 0.101 \\
	$\chi^2$& 1367& 1156 & 343  \\
	dof & 684 & 238 & 431 \\
	\enddata
\end{deluxetable*}	

\section{Image Constrained Light-Curve Models}
\label{sec:newmodel}

Our detection of the exoplanet host star in the high angular resolution follow-up observations can be used to determine the masses and distance of the host star and planet only when the constraints from microlensing light-curve models are also included. One common approach \citep{Bennett2015} is to generate a collection of model parameters that are consistent with the light-curve using MCMC calculations, and then to exclude those models that are inconsistent with the high angular resolution follow-up observation results.

In cases where the modeling includes the microlensing parallax effect ($\pi_{\rm E}$), the fraction of light-curve models in the Markov Chain that are consistent with the high resolution follow-up data can be very small. This is largely because the light-curve measurement of the source-lens relative proper motion constrains the direction of the $\pi_{\rm E}$ vector to be parallel to the relative proper motion. Although $\pi_{\rm E}$ should be present in every microlensing event observed by ground-based telescopes, it is often not included in light-curve modeling analyses since it's parameters cannot be measured very precisely. This approach can be problematic, however, because even when $\pi_{\rm E}$ parameters cannot be measured precisely, they can still influence some of the other model parameters. Therefore, it is prudent to include $\pi_{\rm E}$ values that are consistent with the high angular resolution follow-up data when exploring $\pi_{\rm E}$ models. Furthermore, the event in this paper, OB161195, has a claimed $\pi_{\rm E}$ signal from \textit{Spitzer} observations, so it is helpful to compare the $\pi_E$ signal implied by the high angular resolution follow-up observations to the claim made based on \textit{Spitzer} images.

The vast majority of models from an MCMC run (including $\pi_{\rm E}$ models) are unlikely to be consistent with the follow-up data, which means that very long, and time consuming MCMC runs will be needed to get reasonable sampling of the light-curve models consistent with the data  \citep[e.g.,][]{Bhattacharya2018,Bennett2020}. To address this problem, we have modified our fitting code,  {\fontfamily{lmtt}\selectfont eesunhong\footnote{\url{https://github.com/golmschenk/eesunhong}}} \citep{Bennett1996, Bennett2010model}, to include the constraints from the position and brightness of the lens star identified in the Keck AO images, as described by \citet{Bennett2024}. However, to determine the mass of the host star based on the source-lens relative proper motion (which determines $\theta_{\rm E}$) we need to know the distance to the source star, $D_{\rm S}$, so that the mass-distance relation, equation \ref{mdreltheta}, can be used. This requires us to include $D_{\rm S}$ as a light-curve model parameter, and we use the \citet{Koshimoto2021a} Galactic model to generate a prior $D_{\rm L}$ distribution based on the $t_{\rm E}$ measurement from a more conventional light-curve model without the high angular resolution follow-up observation results. 

Since a companion to the source is likely we tested two light-curve model configurations, a 2L1S (two lens one source) and a 2L2S (two lens two source) with constraints on $\mu_{\rm rel}$ and lens flux from the 2023 Keck data. The best-fit parameters can be found in Table \ref{BondSHVparams}. The 2L2S model improves the fit by $\Delta\chi^2 \sim 15$, however there is no significant difference on the planetary system parameters which are detailed in Section \ref{planetary system parameters}. The table shows the Einstein radius crossing time, $t_{\rm E}$; time of closest approach, $t_0$; the impact parameter, $u_0$; instantaneous projected separation (scaled to $\theta_{\rm E}$), $s$; the angle between the lens axis and the source trajectory, $\alpha$; source star crossing time, $t_*$; the mass ratio, $q$; the Einstein angular ring radius, $\theta_{\rm E}$; and the source color and magnitude. 
	
Since we introduce a source companion to the primary source, additional parameters need to be included, such as the time of closest approach ($ t_{0,2}$) and impact parameter ($u_{0,2}$) of the second source, and the source 2 to source 1 flux ratios in $I$ and $V$ (f2rI and f2rV, respectively). 

The two source stars are unlikely to have the same velocity due to orbital motion, therefore, their $t_{\rm E}$ and $\alpha$ values will be different. 
The difference in these values can be described by $\rm dt_{\rm E21}$ implied by $t_{\rm E2} - t_{\rm E1}$, and $\rm d\alpha$ implied by $\alpha_{\rm s2} - \alpha_{\rm s1}$, where $t_{\rm E1} = t_{\rm E}$ and $\alpha_{\rm s1} = \alpha$ \citep{Bennett2010apr, Bennett2018b}. These parapeters describe a circular orbit with no constraint on orbit orientation.

\begin{figure}[h!]
	\centering
	\includegraphics[width=8.5cm]{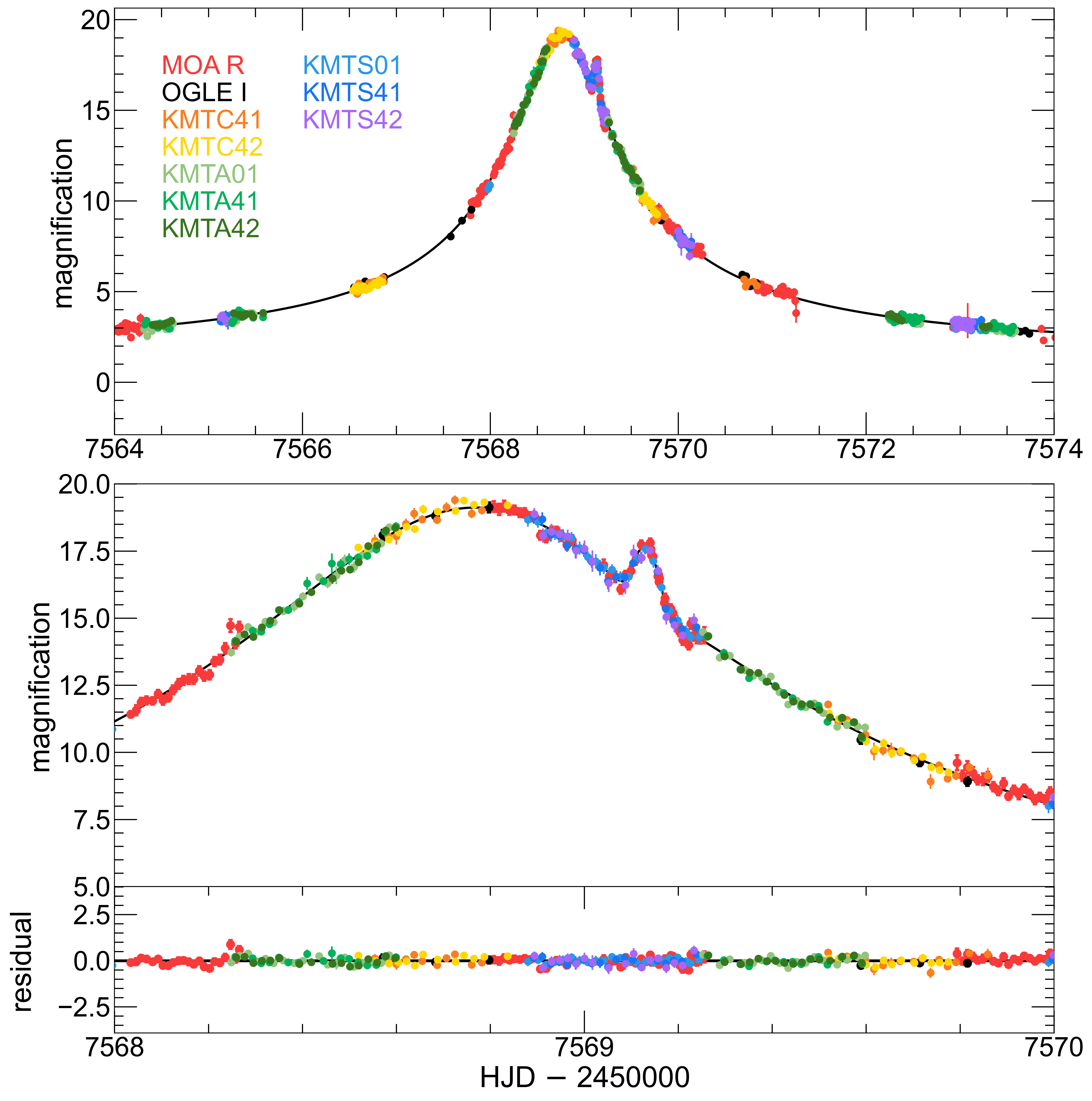}
	\caption{ The OB161195 best fit 2L2S light-curve with constraints from the high angular resolution Keck data - using the full data set (MOA, OGLE and KMTNET).}
	\label{fig:model}
\end{figure}

 For each model configuration, 2L1S and 2L2S, we find a close and wide fit with a difference of $\Delta\chi^2 \sim 1$. For our analysis we continue with the 2L2S model since the $\chi^2$ is slightly favored and we know from our Keck data that there is excess flux at the location of the source. The remodeled light-curve using the full dataset (MOA, OGLE and KMTNet) is shown in Figure \ref{fig:model}.

We note that an alternative binary source model was considered since there is a companion of the source, and a short term planetary perturbation can be mimicked by a binary source system with an extreme flux ratio \citep{Gaudi1998}. This involved the high magnification microlensing of a very faint source companion, which would have been undetectable in our Keck images. Such models are excluded by the light-curve data, but we did find a model, with a more extreme flux ratio of 0.0023, that was a better fit to the MOA and OGLE data than the extreme flux ratio model presented by Bond et al. (2017). However, this model is excluded more strongly by the combined  MOA, OGLE, and KMTNet data with a $\Delta\chi^2 > 110$.

\subsection{{Source colors and CMD}}

\begin{figure}[t!]
	\centering
	\includegraphics[width=9cm]{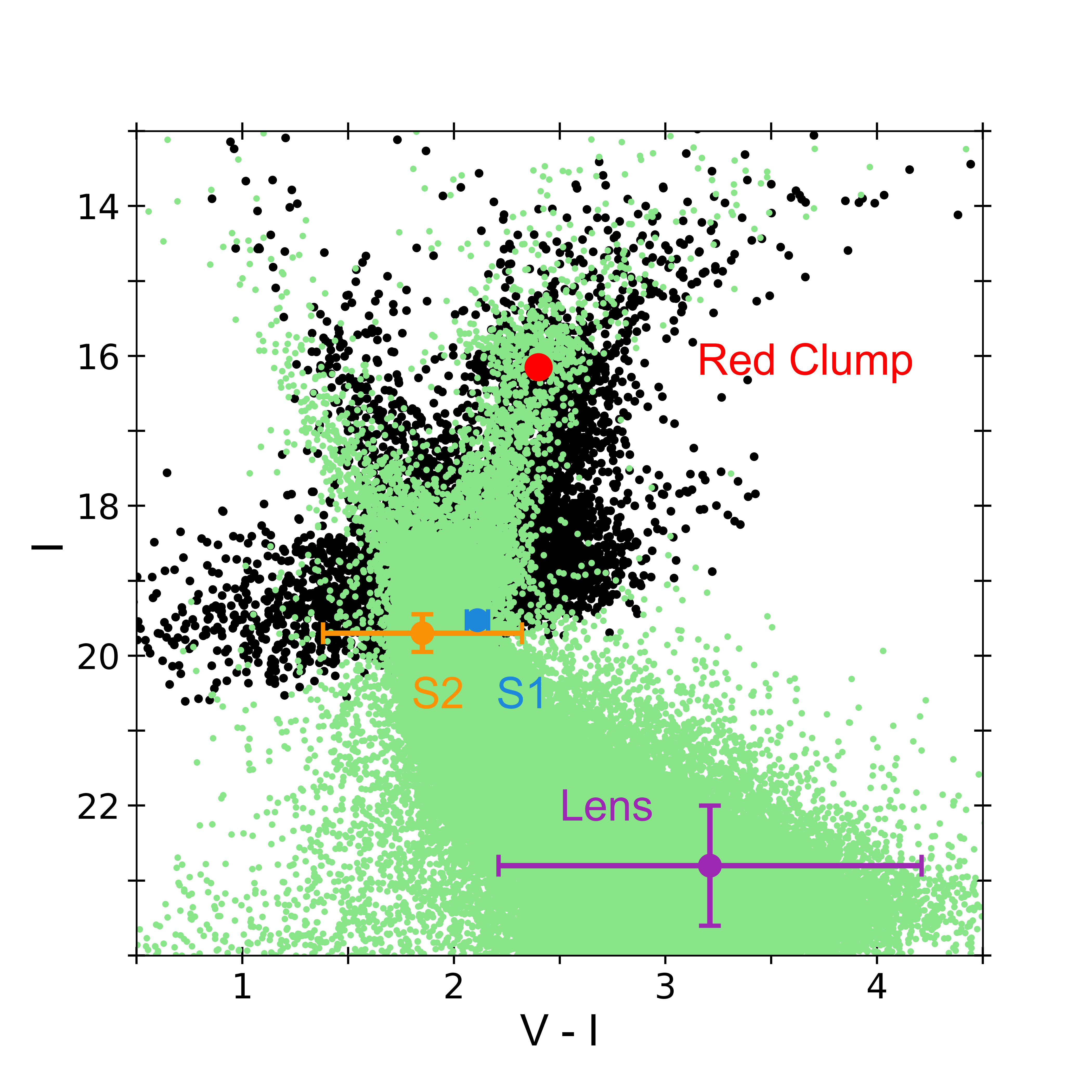}
	\caption{Color-magnitude diagram (CMD) for the OB161195 field. In black are the OGLE-III stars within 90 arcseconds of the target, and with the HST CMD of the Stanek window in green \citep{Stanek1994, Terry2020}. The red spot is the Red Clump centroid, the orange and blue are source 2 and source 1, respectively. In purple is the location of the lens. }
	\label{fig:CMD}
\end{figure}

We plot the positions of both sources and lens on the CMD in Figure \ref{fig:CMD}. The MCMC distributions for the source's color and magnitude from the 2L2S remodeled light-curve are:
\begin{gather*}
	(V-I, I )_{\rm s1}= (2.111, 19.525) \pm (0.053, 0.031),\\
	(V-I, I )_{\rm s2}= (1.851, 19.697) \pm (0.469, 0.251).
\end{gather*}

Using extinction values of $A_{\rm I} = 1.762 \pm 0.018$ and $E(V-I) = 1.408 \pm 0.007$ \citep{Nishiyama2006, Surot2019} we find the dereddened magnitudes: 
\begin{gather*}
	(V-I, I )_{\rm s1,0}= (0.703, 17.763) \pm (0.053, 0.036),\\
	(V-I, I )_{\rm s2,0}= (0.443, 17.935) \pm (0.469, 0.252).
\end{gather*}

The dereddened magnitudes can be used to calculate the angular source radius, $\theta_{*}$, using the following relation from \cite{Boyajian_2014}: 
\begin{equation}
	\log_{10}[2\theta_{*}/(1\rm mas)] = 0.5014 + 0.4197(V-I)_{\rm s0} - 0.2I_{\rm s0}.
\end{equation}

Therefore, we find $\theta_{*} = 0.877 \pm 0.054\ \mu\rm as$ for the primary source in the 2L2S model. We can estimate the relative proper motion in the geocentric reference frame from the light-curve parameters using $\mu_{\rm rel, G} = \theta_{*}/t_{*} = 10.23 \pm 0.77$ mas/yr. This value cannot be directly compared to our measured values from the Keck images, since those are in the heliocentric reference frame. Further details about converting from heliocentric to geocentric frames are given in \ref{planetary system parameters}.

The MCMC distribution of the lens $V$ and $I$ measurements from the remodeled 2L2S light-curve are $(V-I, I )_{\rm L}= (3.211, 22.803) \pm (1.265, 0.809)$ mag. For the source distance we find $D_{\rm S} = 8.74 \pm 0.63$ kpc. The source distance for the 2L2S model can be used to estimate the physical separation (in AU) between the two sources when multiplied by the separation in arcsecond units ($S1-S2 = 6.318 \pm 1.737$ mas). Therefore, we calculate the physical separation between the two sources to be $\sim 53 \pm 15$ AU.

Although the physical separation between the two sources is such that it can barely be detected in the light-curve model, it can be estimated from the length of the separation vector between the two sources multiplied by  $\theta_{\rm E}$ and $D_{\rm S}$, thus

\begin{equation}
	\sqrt{\left(\frac{t_{0,2}-t_0}{t_{\rm E}}\right)^2 + (u_{0,2} - u_0)^2} \times \theta_{\rm E} D_{\rm S}.
\end{equation}

This calculation gives a separation of 7.017 $\pm$ 1.895 AU at the peak of the microlensing event. If we assume the relative tranverse velocity of the two source stars to be $\sim 10$ km/sec = 2 AU/year, this would imply the two sources could have separated an additional $\sim$ 14 AU at the time of the follow-up observations in 2023. Even though this is not in full agreement with the value we find in the 2023 Keck data, it should be noted that the positions of the two sources in the Keck data are not well constrained (due to their small separation which is $<$ 1 pixel = 9.942 mas), and the uncertainty on their separation is likely underestimated. However, this does not affect the physical parameters of the planetary system. 

\section{Planetary System Parameters}
\label{planetary system parameters}

%{\color{red} We proceed using the lightcurve model parameters for the 2L2S best fit model.}

Without additional constraints (such as the lens flux from Keck) the physical parameters of a lensing system can be derived by combining two second order effects on the microlensing light-curve. These are finite source effects and the microlensing parallax effect. The former provides a measurement for the source radius crossing time, $t_*$, which can be used to determine the angular Einstein radius, $\theta_{\rm E}$. Therefore, a mass-distance relation can be obtained. 

\begin{equation}
	\label{mdreltheta}
	M_{\rm L} = \frac{\theta_{\rm E}^2}{\kappa \pi_{\rm rel}},\ 
	\pi_{\rm rel} = AU\left(\frac{1}{D_{\rm L}} - \frac{1}{D_{\rm S}}\right),
\end{equation}
where $\kappa = 8.144\ \rm mas\ M_\odot^{-1}$. This equation has two unknowns however; lens mass and distance. If a microlensing parallax effect, $\pi_{\rm E}$, is measured from the light-curve, then an additional mass-distance relation can be obtained \citep{Gould1992A}, which when combined with Equation \ref{mdreltheta}, yields: 

\begin{figure}[h!]
	\includegraphics[width=9.5cm]{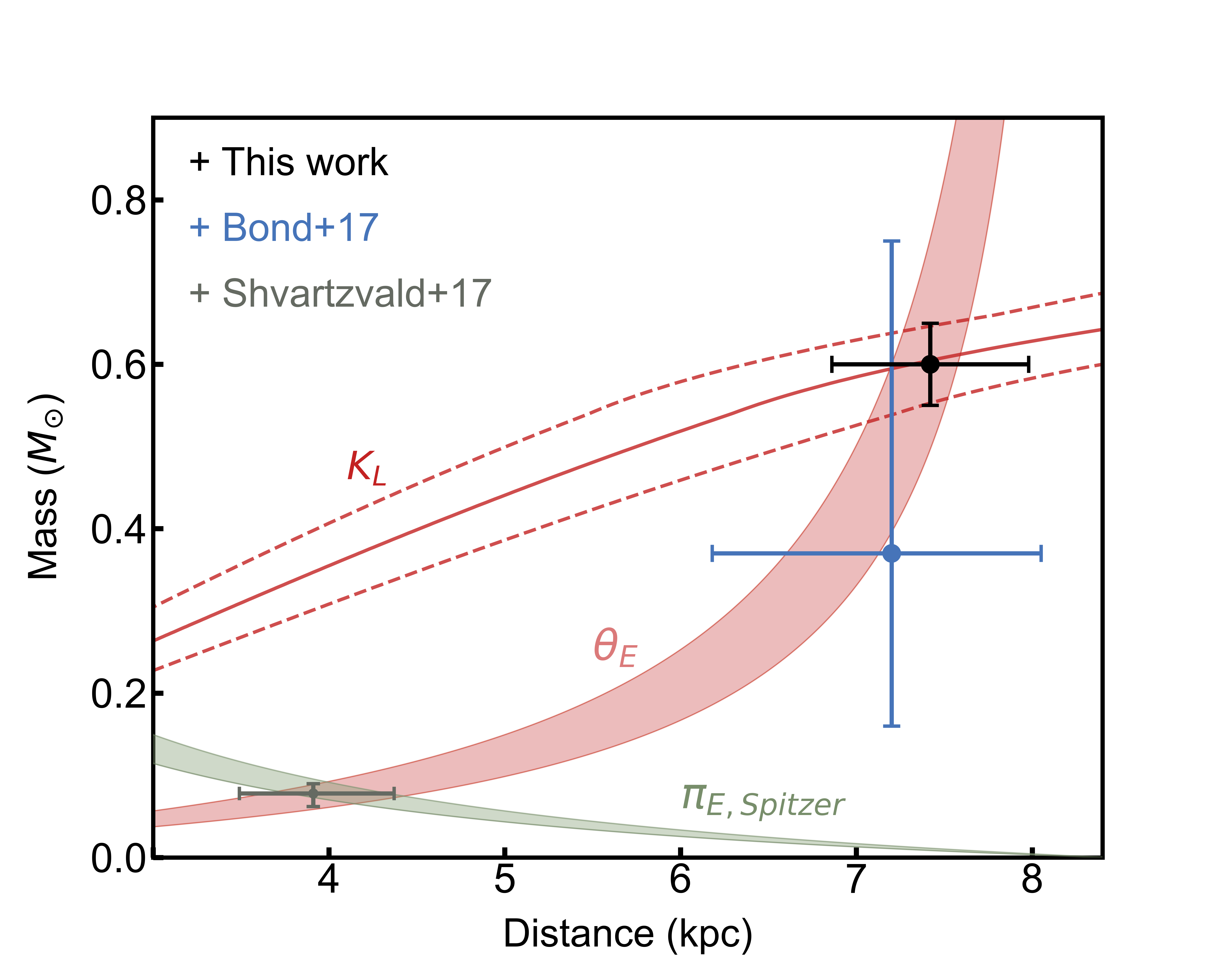}
	\caption{ Mass-distance relations for OGLE-2016-BLG-1195, obtained using the Einstein ring radius ($\theta_{\rm E}$) and parallax ($\pi_{\rm E}$) from remodeling the light-curve with the measured 2023 Keck $\mu_{\rm rel}$ constraint, and also the 2023 Keck K-band flux constraint ($K_{\rm L}$). The planet-host properties are derived from the intersection of the two most reliable mass-distance relations. The host system properties estimated by \citetalias{Bond2017} and \citetalias{Shvartzvald2017} are shown as crosses in blue and green, respectively. The Spitzer parallax constraint obtained by \citetalias{Shvartzvald2017} is shown in gray. This is a simplified presentation of the results that include errors measured on the host star magnitudes and average errors on the $\theta_{\rm E}$ values.}
	\label{fig:1195massdistance}
\end{figure}

\begin{figure*}[t!]
	\centering
	\includegraphics[width=16cm]{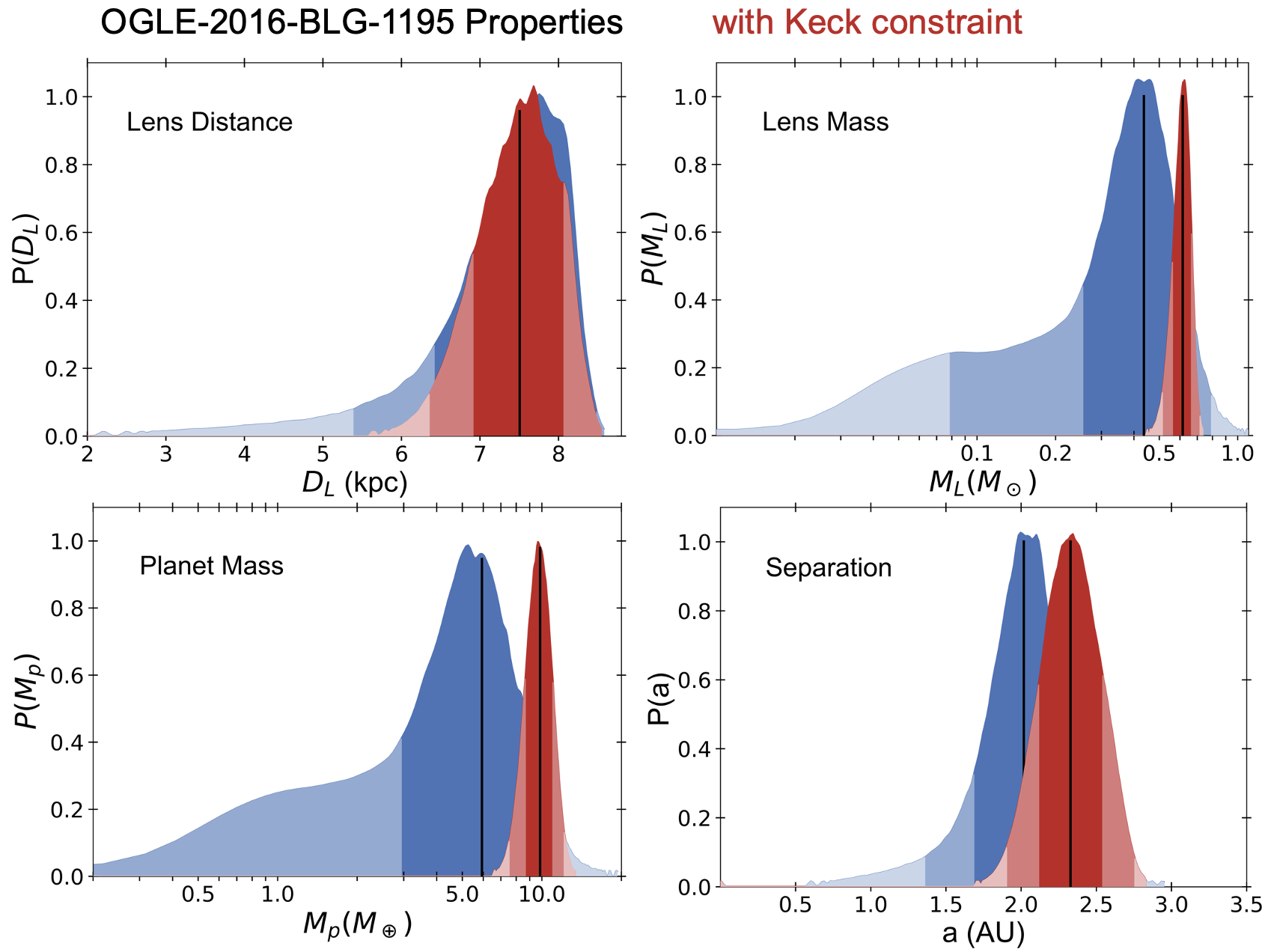}
	\caption{Bayesian posterior probability distributions for the host and planetary companion mass, their separation, and lens system distance. Using only light-curve constraints are shown in blue, and with additional constraints from the follow-up Keck data in red. The center of the distributions (68.3$\%$) are shaded in dark red and blue. The remaining 95.4 $\%$ are shaded with lighter hues. The dark vertical line indicates the median of the probability distribution of each parameter.}
	\label{fig:histo}
\end{figure*}

\begin{equation}
M_{\rm L} = \frac{\pi_{\rm rel}}{\kappa \pi_{\rm E}^2} = \frac{\theta_{\rm E}}{\kappa \pi_{\rm E}}.
\end{equation}

Many microlensing events do not have a measurable parallax effect though, and at best, only a projection of the parallax vector parallel to the Earth's orbit is measured, leaving room for degeneracies. Thus, additional constraints are needed for accurate estimates of the physical parameters to be made, such as high angular resolution follow-up observations with Keck. These can be implemented from an empirical mass-luminosity relation with the measured Keck apparent magnitude. The empirical mass-luminosity relations (also used in \cite{Bennett2015, Bennett_2016, Bennett2018}) are from \cite{Henry1993} ($M_{\rm L} \geqslant 0.66\ M_{\odot}$), \cite{Delfosse2000} ($0.54\ M_{\odot} \geqslant M_{\rm L} \geqslant 0.12\ M_{\odot}$) and \cite{Henry1999} ($0.10\ M_{\odot} \geqslant M_{\rm L} \geqslant 0.07\ M_{\odot}$). The empirical mass-luminosity relation merges these three relations, smoothly interpolating between them. 

The foreground extinction is also considered for the mass-luminosity relation. Following \cite{Drimmel2001} we use the following equation to estimate the foreground extinction for the lens:

\begin{equation}
	A_{i,L}=\frac{1-e^{-|D_{L}({\rm sin}\ b)/h_{\rm dust}}|}{1-e^{-|D_{S}({\rm sin}\ b)/h_{\rm dust}}|}\ A_{i,S}\ ,
\end{equation}

where $h_{\rm dust}$ is the dust scale height, $b$ is the Galactic latitude of the target ($b = -2.475$), and $i$ is an index refering to the passband.

In addition, since Keck can measure the source-lens proper motion, $\theta_{\rm E}$ can be recalculated using $\theta_{\rm E}=t_{\rm E} \mu_{\rm rel}$. However, the microlensing light-curve model is in the geocentric reference frame, so we must convert our heliocentric relative proper motion, using:

\begin{equation}
	\label{projected_velocity}
	\mu_{\rm rel, G} = \mu_{\rm rel, Hel} - \frac{v_{\oplus} \pi_{\rm rel}}{\rm AU} ,
\end{equation}

where $v_{\oplus}$ is the projected velocity of the earth relative to the sun at the time of peak magnification. For OB161195, this was $v_{\oplus E, N} = (28.861, -0.991)\ \rm km/sec = (6.085, -0.209)\ \rm AU/year$ at HJD'= 7569. With $\pi_{\rm rel}$ as the relative parallax, we can rewrite equation \ref{projected_velocity} as:

\begin{equation}
	\mu_{\rm rel, G} = \mu_{\rm rel, Hel} - (6.085, -0.209) \times \left(\frac{1}{D_{\rm L}} - \frac{1}{D_{\rm S}}\right).
\end{equation}

$\mu_{\rm rel, G} $ and $\mu_{\rm rel, Hel}$ are in units of mas/year, and $D_{\rm L}$ and $D_{\rm S}$ are in units of kpc. This relation is used in the Bayesian analysis of the light-curve with the high resolution follow-up constraints from Keck. Therefore, we find $\mu_{\rm rel, G} = 10.91 \pm 0.26\ $ mas/yr and  $\theta_{\rm E}=0.290 \pm 0.013$ mas. These values are the weighted sum of the values from the Keck images and the light-curve parameters ($t_{*}$ and $\theta_{*}$). This is within $1\sigma$ agreement with $\mu_{\rm rel, G} = \theta_{*}/t_{*} = 10.23 \pm 0.77$ mas/yr, calculated previously.
 
 For OB161195 we combine constraints from the empirical mass-luminosity relation and Equation \ref{mdreltheta}, plotting them on a mass-distance diagram shown in Figure \ref{fig:1195massdistance}. These two relations are shown in different shades of red, with their intersection (black cross) indicating the result we find by summing over the MCMC runs of the 2L2S model using a Galactic model (discussed in section \ref{sec:newmodel}). It should be noted that Figure \ref{fig:1195massdistance} is a simplified presentation of the results, whereas Figure \ref{fig:histo} and the results shown in Table \ref{finalparams} include the effects of the different correlations with other model parameters from the MCMC analysis.

We run two Galactic models with different planet hosting priors (M$^{\wedge}0$ and M$^{\wedge}1$) and sum over the MCMC results. We compare the results of both priors, finding that the mass measurement does not vary significantly between the two. The M$^{\wedge}0$ prior is most commonly used in Bayesian analyses of microlensing events that lack mass measurements \citep{Bennett2014}. However, more recent studies (\cite{Bennett2024}, \cite{Koshimoto2021a}) indicate that the probability of hosting a planet with a fixed mass ratio increases linearly with the host mass, i.e. M$^{\wedge}1$ model. For OB161195 the final values that we use are from the M$^{\wedge}1$ model. However, because we have a mass estimate from the Keck follow-up data, there is no significant difference between the two models. 

These results are presented in Table \ref{finalparams} and Figure \ref{fig:histo}, where the blue histograms indicate the parameters without Keck constraints, and red histograms with Keck constraints (these are generated using the model with the M$^{\wedge}1$ prior). Therefore, we find that the OB161195 system is a cold sub-Neptune with a mass of $M_{\rm p} = 10.08\pm1.18\ M_{\oplus}$ orbiting a star with $M_{\rm L}=0.62\pm0.05\ M_{\odot}$, at a distance of $D_{\rm L}=7.45\pm0.55$ kpc towards the Galactic Bulge. The projected separation between planet and host is $r_\perp=2.24\pm0.21$ AU, placing it beyond the snow line. 

In Figure \ref{fig:1195massdistance} we can see a graphical summary of the various conclusions for OB161195, where \citetalias{Bond2017} is represented with a blue cross, \citetalias{Shvartzvald2017} with gray and for this work, black. We immediately notice the drastic difference that the added parallax measurement from \textit{Spitzer} has made to the physical parameters. In general, parallax measurements can be difficult to constrain since they require adequate sampling and event coverage for their effects to be seen on a light-curve. Even parallax measurements obtained with space based telescopes, such as \textit{Spitzer}, can be ambiguous if there are systematic errors in the photometry that are difficult to detrend due to poor cadence and signal coverage. In \citetalias{Shvartzvald2017} the \textit{Spitzer} light-curve data have low signal-to-noise because the target is faint and covers only the trailing end. In addition, it could be affected by low level systematic errors in the photometry \citep{Koshimoto2020}, and therefore it is necessary to investigate the \textit{Spitzer} data further.

\begin{deluxetable}{ l c c r r }
	\label{finalparams}
	\tablecaption{Physical parameters from \citetalias{Bond2017} and \citetalias{Shvartzvald2017} compared to those found in this paper from the best fit 2L2S MCMC values from both planet hosting priors, M$^{\wedge}0$ and M$^{\wedge}1$. We adopt the  M$^{\wedge}1$ values. \label{masses} }
	\tablewidth{0pt}
	\tablehead{  & \colhead{M$^{\wedge}0$}  & \colhead{M$^{\wedge}1$} & \colhead{\citetalias{Bond2017}}&\colhead{\citetalias{Shvartzvald2017}} 
	 }
	 % end header.
	%\hline
	%\hline
	\startdata
	$M_{\rm L}$ ($M_\odot$)& 0.62 $\pm$ 0.05 & 0.62 $\pm$ 0.05 & $0.37^{+0.38}_{-0.21}$ & $0.078^{+0.016}_{-0.012}$  \\   
	$M_{\rm p}$ ($M_\oplus$) & 10.04$\pm$ 1.18&  10.08 $\pm$ 1.18 & $5.10^{+5.50}_{-2.85}$ & $1.43^{+0.45}_{-0.32}$ \\
	$D_{\rm L}$ (kpc) & 7.42 $\pm$ 0.56 & 7.45 $\pm$ 0.55 & $7.20^{+0.80}_{-1.02}$  & $3.19^{+0.42}_{-0.46}$ \\
	$r_\perp$ (AU) & 2.23 $\pm$ 0.21& 2.24$\pm$ 0.21 & $2.01^{+0.27}_{-0.32}$ & $1.16^{+0.16}_{-0.13}$\\	
	\enddata
\end{deluxetable}

\begin{figure}[h!]
	\centering
	\includegraphics[width=9.5cm]{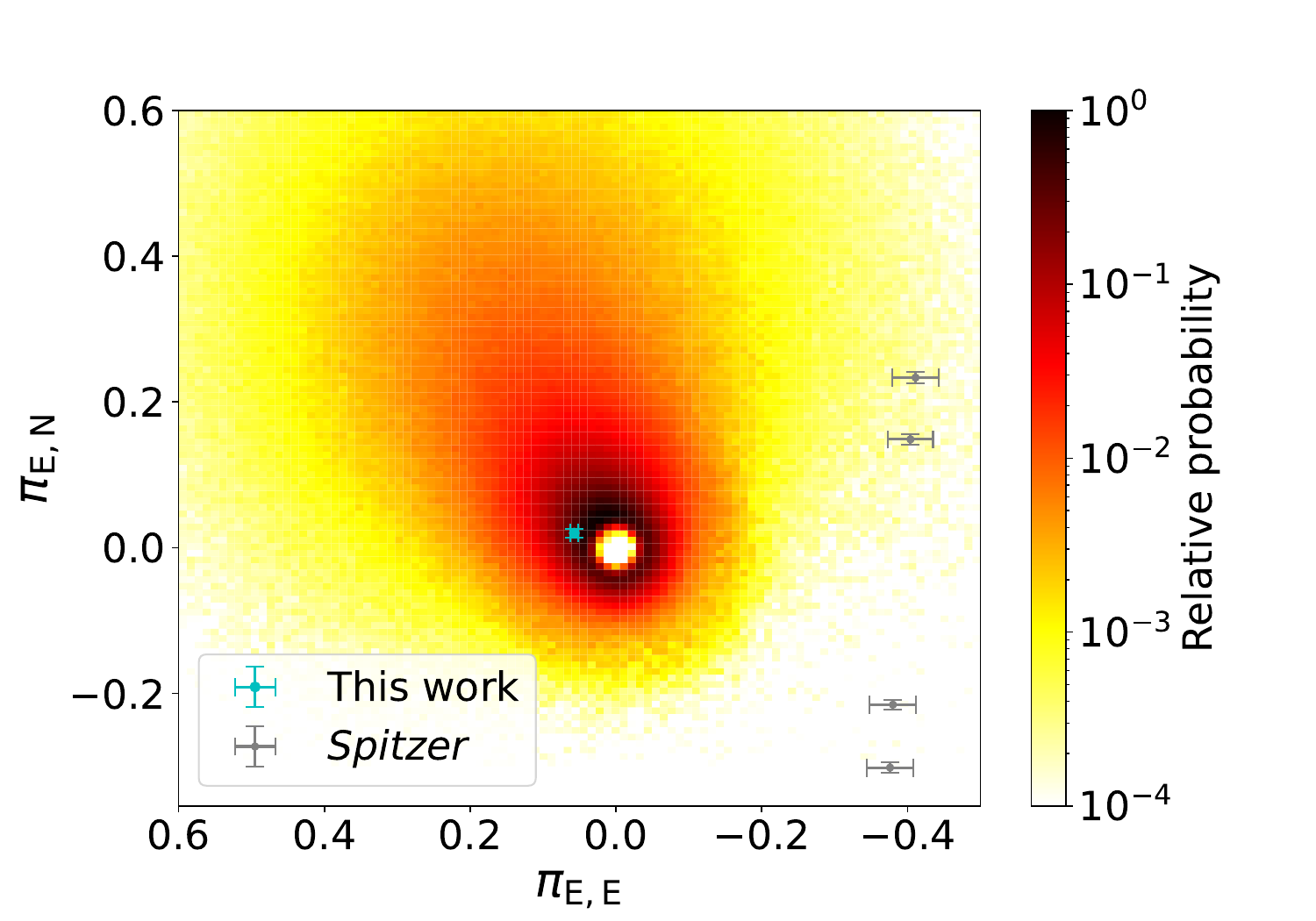}
	\caption{Two-dimensional parallax distribution for the lens in OB161195. The blue cross indicates the value obtained from this work, with constraints from Keck, and the gray crosses are the values presented in \citetalias{Shvartzvald2017} using the \textit{Spitzer} dataset. The color-scale shows the relative probability based on a Galactic model}
	\label{fig:Pi2D}
\end{figure}

\subsection{Re-analysis of the \textit{Spitzer} data}
\label{Spitzer}

The large discrepancy between \citetalias{Shvartzvald2017}'s constraints on the planetary system's physical parameters and those reported by \citetalias{Bond2017} and this work, can be explained by the microlensing parallax direction claimed in \citetalias{Shvartzvald2017}. The value they present, $\pi_{\rm E,N}, \pi_{\rm E,E} = (-0.3016, -0.3770) \pm ( 0.0074, 0.032)$, is  highly unlikely when compared to the stellar population of the Galactic disk (see Figure \ref{fig:Pi2D}). In fact all their Spitzer solutions yield unlikely parallax values, such as $\pi_{\rm E,N}, \pi_{\rm E,E} = (-0.2158, -3820) \pm ( 0.0066, 0.032)$, $\pi_{\rm E,N}, \pi_{\rm E,E} = (0.2350, -0.413) \pm \newline ( 0.008, 0.031)$, and $\pi_{\rm E,N}, \pi_{\rm E,E} = (0.1491, -0.404) \pm ( 0.0075, 0.032)$. In \citetalias{Shvartzvald2017}, they explain that the unlikely parallax direction (and therefore a counter-rotating lens) is a direct result of the magnification seen by \textit{Spitzer}, and that high resolution imaging can resolve this issue. Our high resolution Keck observations indicate that the apparent Spitzer parallax signals are not real. Using constraints from Keck the measured parallax direction is $\pi_{\rm E,N}, \pi_{\rm E,E} = (0.0194, 0.0551) \pm ( 0.0057, 0.0052)$. Counter-rotating low-mass stars (such as brown dwarfs) detections have been claimed before with the microlensing method (e.g. OGLE-2017-BLG-0896 \cite{Shvartzvald2019} and MOA-2015-BLG-231 \cite{Chun2019}). Both these events however, have large parallax measurements from \textit{Spitzer}, which have not yet been validated with high angular resolution follow-up observations. Counter rotating brown-dwarf solutions are likely evidence for systematic errors.

Therefore, we decided to re-assess the \textit{Spitzer} light-curve data of OB161195. Observations of this event were obtained on two occasions with \textit{Spitzer}/IRAC \citep{Fazio2004} in the 3.6 $\mu$m channel as part of the 2016 \textit{Spitzer} microlensing campaign during July 12--24, 2016 at a cadence of one epoch per day. This corresponds to a total of 17 epochs with 6 dithered frames each. Additionally, 7 epochs of baseline observations of OB161195 were collected during July 12--24, 2019 at the same cadence. 

\begin{figure*}
	\centering
	\includegraphics[width=\linewidth]{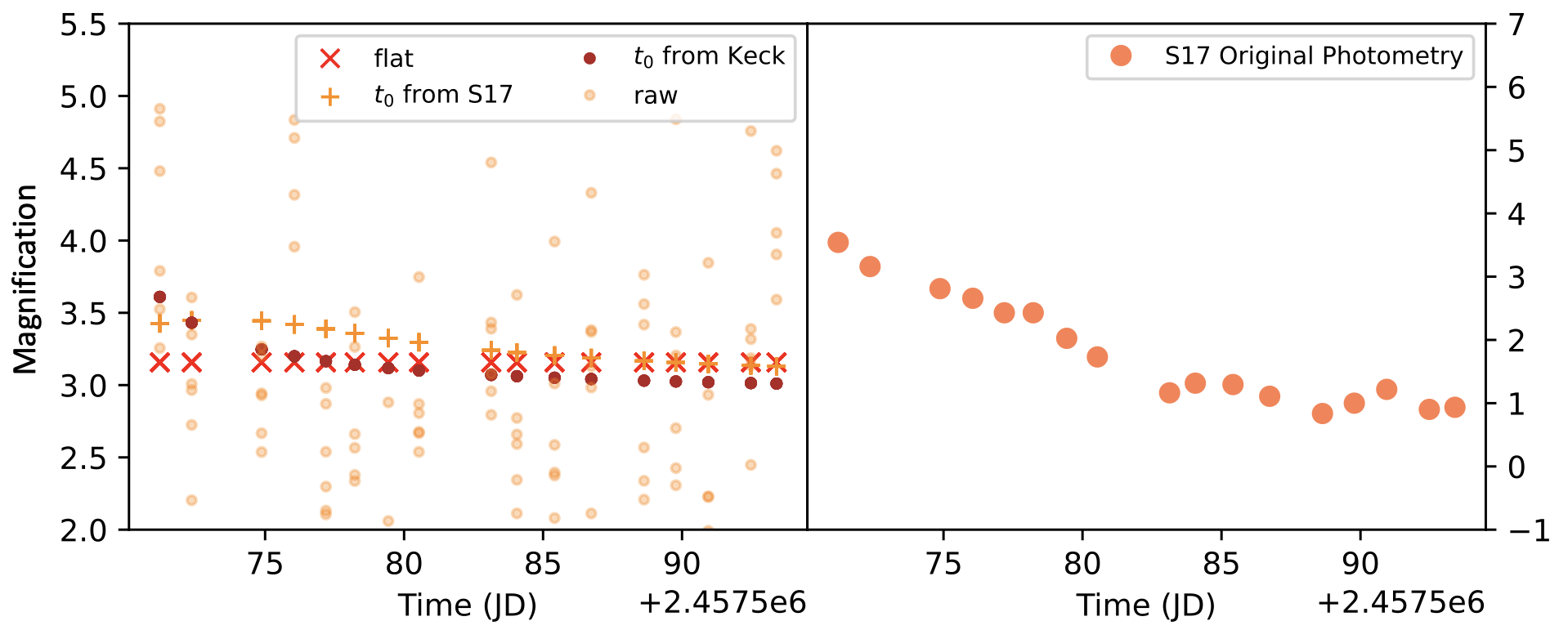}
	\caption{Left Panel: Three different models of the \textit{Spitzer} data. The flat-line model from the 2019 baseline \textit{Spitzer} data has a $\chi^2 = 120.059$. The $t_0$ model inferred from \citetalias{Shvartzvald2017} has a $\chi^2 = 121.263$. The $t_0$ model inferred from the Keck follow-up results has a $\chi^2 = 102.117$. Right Panel: \citetalias{Shvartzvald2017}'s original magnification curve from \textit{Spitzer} photometry as a comparison, with magnification ranging from $-1 < y < 7$. }
	\label{fig:PLD}
\end{figure*}

Following \cite{Dang2020}, we perform aperture photometry then model the detector systematics using Pixel Level Decorrelation (PLD). We first identify the 6 dithered regions of the 256$\times$256 pixel detector where the target is positions and perform aperture photometry by using a 3$\times$3 square aperture. We then calculate the relative flux measured by each pixel and use them as regressor for our PLD noise models. We define the measured flux as a combination of the astrophysical model and PLD systematics model as a multiplactive term. Assuming a 1S1L model for the astrophysical model, we use an MCMC to simultaneously evaluate the astrophysical parameters and PLD coefficients. We find that with such a small number of epochs, the detector systematics at each dither position are poorly sampled, so a self-calibration approach such as the PLD tends to overfit the data as suggested by the overly optimistic residuals. Consequently, the PLD analysis is inconclusive, which motivates a second look at the \textit{Spitzer} frames. Nonetheless, we can compare our raw photometry with that presented in \citetalias{Shvartzvald2017} which uses \cite{CalchiNovati2015}'s PSF fitting procedure.

After close inspection of our initial photometry, there is no apparent downward trend as reported by \citetalias{Shvartzvald2017}, which suggests that the magnification of the source is not significantly greater than the instrumental noise and the potential contamination from nearby sources. We compared the original 2016 observations with the 2019 baseline data and we note that the scatter between dithers in 2016 is systematically larger than that observed in the 2019 observations. This could be indicative that the 2016 observations either exhibit stronger instrumental systematics or that the time-variable contamination from nearby sources was more prominent. Hence, given that the source is faint and that the \textit{Spitzer} magnification could be as large as the instrumental noise, if not modeled properly, we suggest that the discrepancy between the physical parameters of OB161195 inferred in this work and the results reported by \citetalias{Shvartzvald2017} could be due to a poor characterization of the instrumental systematics of this particular event. 

While PLD tends to overfit OB161195's Spitzer observations, we investigate the discrepancy physical parameters inferred from \citetalias{Shvartzvald2017} and our new Keck constrains by fitting 3 different light-curve models to our raw photometry with a PLD correction: 1) a flat line , a single-lens model with $t_0$ fixed to that reported by \citetalias{Shvartzvald2017}, and 3) a single-lens model with $t_0$ fixed to that predicted from our Keck constrains. These fits can be seen in Figure \ref{fig:PLD}. In this analysis, we find that the model with $t_0$ from \citetalias{Shvartzvald2017} is disfavored, with the highest $\chi^2$ compared to the other models. The model that is favored when comparing $\chi^2$ is that from the Keck follow-up results, although the difference is small and not significant enough to conclude which of the two is better. Therefore, our ultimate deduction is that the \textit{Spitzer} dataset is inconclusive, and the parallax yielded from such sparcely sampled data is not a reliable avenue for a mass-measurement to be made.

%{\color{red} Add and image showing Spitzer fits file compared to Osiris fits file - JP}.

\section{Discussion and Conclusion}
\label{sec:conclude}

Our 2023 high angular resolution follow-up observations of the microlensing event OB161195 have allowed us to resolve the source and the lens with a separation of $\sim$ 79 mas, 7.12 years after the peak of the event. By directly measuring the lens' K-band flux, and the amplitude and direction of the source-lens relative proper motion we have estimated the mass and distance of this planetary system. We explore a scenario where the source star has a companion since we find excess flux at the location of the source, and we rerun the microlensing light-curve model with a 2L2S configuration which improves the model by $\Delta\chi^2 \sim 15$. We combine this with a 3-star fit on the 2023 Keck data. Microlensing events with a binary source star are not uncommon, e.g. \citet{Bennett2018b, Han2023, Han2023a}. Effects on the light-curve due to a binary source can be seen in the form of xallarap (source acceleration due to due to a binary companion) \citep{Dominik1998},  or magnification of the second source \citep{Griest1992}.
	
	%{\color{red} this is because even though binary sources can have an effect on the light curve - in this case the separation between the 2 source is too large ($\sim$6 mas or $\sim$50 AU) for any xallarap effect to be seen.}}

We find that OB161195 is not an ultracool dwarf orbited by an Earth-mass planet, instead we find that it is a cold sub-Neptune (or super-Earth) orbiting an M-dwarf, supporting the conclusion of \citetalias{Bond2017} which was result was reached using a Bayesian analysis with a standard Galactic model and a prior that assumes that for a mass ratio, $q$, the probability of hosting a planet is independent of the host mass. This prior assumption that is commonly used in microlensing has to be updated as described in Section \ref{sec:newmodel}. New research (\citet{Bennett2024, Koshimoto2021a}) indicates that planet detection frequency scales linearly with host mass, for a fixed mass ratio. Although the two priors differ (M$^{\wedge}$0 and M$^{\wedge}$1), the final results for OB161195  are completely independent of the prior because we have enough measurements to constrain the masses. Cases that only have $\theta_{\rm E}$ measurements (such as \citetalias{Bond2017}) are sensitive to the prior assumption. 

Although OB161195 is not an ultracool dwarf as reported in \citet{Shvartzvald2017}, false detections of these substellar objects could have negative implications on the understanding of their formation and evolution. Ultracool dwarfs are predicted to make up $15-30\%$ of stars in the Galaxy \citep{Kroupa2001, Muzic}, however their formation and evolution are currently only incompletely understood, since direct mass measurements are difficult to achieve for these faint objects. Techniques such as transits (e.g \cite{Carmichael2020}) and direct imaging (e.g \cite{Burgasser_2006}) have had some success with ultracool dwarf binary systems, however, these methods are only sensitive to close binaries with mass ratios $q\gtrsim0.4$. Additionally, they can only measure the total mass of the system, not the individual masses. Microlensing complements these techniques as it has the ability to detect wide-orbit binaries with low mass ratios. The individual masses can also be measured if the microlensing light-curve exhibits both finite source effects and a parallax effect. With OB161195 as an example, however, parallax effects measured by \textit{Spitzer} are often contaminated by systematics and/or have low signal-to-noise \citep{Koshimoto2020}. Therefore, high resolution follow-up observations are the optimum solution to determine the physical properties of these low-mass ratio systems.

Parallaxes in general are not easy to measure and typically result in degenerate light-curves. In the case of \textit{Spitzer} parallaxes, \cite{Koshimoto2020} found that $>75\%$ of parallaxes obtained by the \textit{Spitzer} campaign were higher than the median averages predicted by Galactic models, from a study of 50 microlens parallax measurements. They concluded that the main source of this discrepancy came from systematic errors in the \textit{Spitzer} photometry. In particular \cite{Koshimoto2020} draw attention to this event, noting that the \textit{Spitzer} light-curve does not show a clear peak feature and attempts to measure a faint source star, both of which contribute to unusually high systematic errors. 

A point of interest for OB161195 is its very low mass-ratio, $q= (4.94 \pm 0.49) \times 10^{-5}$, because it puts it just below the estimated power law break seen in the statistical analyses of $\sim$30 microlensing planetary events by \cite{Suzuki2016, Suzuki2018}. In their work they studied the measured mass ratios of events detected from 2007 to 2012 by the MOA survey, and compared them to the planet distributions from core accretion theory population synthesis models. Although the core accretion models predict planets with mass ratios $q < 10^{-5}$, measurements indicate a break at $q \sim 1.7\times10^{-4}$, indicating that we do not know the occurrence rates of planets beyond this. The precise location of the break is not well constrained, with its location more likely to be at $q = 6.7^{+9.0}_{-1.8} \times 10^{-5}$. Therefore, OB161195 with $q\approx4.84 \times 10^{-5}$, is just below the mass ratio function break. 

High resolution follow-up observations are crucial for accurate mass measurements of microlensing systems. In this paper we have presented the event OB161195 which had contradictory conclusions from the two detections papers, \citetalias{Bond2017} and \citetalias{Shvartzvald2017}. Through multi-epoch high-resolution Keck AO data we have shown that \textit{Spitzer} parallax measurements are not always a reliable mass measurement method, as they are prone to systematic errors in the photometric data when the target is faint. Low cadence observations from the space telescope can also impact the credibility of the parallax measurement. The types of planets and host stars that microlensing can detect, complement the other techniques, offering a unique window into the population and distribution of these objects in our Galaxy.

	\acknowledgments
	
		\footnotesize{This work is supported by NASA through grant NASA-80NSSC18K0274 and by the University of Tasmania through the UTAS Foundation and the endowed Warren Chair in Astronomy and the ANR COLD-WORLDS (ANR-18-CE31-0002. This research was also supported in part by the Australian Government through the Australian Research Council Discovery Program (project number 200101909) grant awarded to AC and JPB. The Keck Telescope observations and analysis were supported by a NASA Keck PI Data Award, administered by the NASA Exoplanet Science Institute. Data presented herein were obtained at the W. M. Keck Observatory from telescope time allocated to NASA through the agency’s scientific partnership with the California Institute of Technology and the University of California. The Keck I and Keck II real-time controller upgrades are funded by the National Science Foundation’s Mid-Scale Innovations Program award AST-1836016 and Major Research for Instrumentation Program award AST-1727071, respectively. The Observatory was made possible by the generous financial support of the W. M. Keck Foundation. The authors wish to recognize and acknowledge the very significant cultural role and reverence that the summit of Mauna Kea has always had within the indigenous Hawaiian community. We are most fortunate to have the opportunity to conduct observations from this mountain. DPB and AB were also supported by NASA through grant NASA-80NSSC18K0274. Work by C.R. was supported by the Alexander von Humboldt Foundation. L.D. acknowledges support in part through the Technologies for Exo-Planetary Science (TEPS) PhD Fellowship, the Natural Sciences and Engineering Research Council of Canada (NSERC)'s Canada Graduate Scholarship and Banting Postdoctoral Fellowship.  LD, TBK, and NBC acknowledge camaraderie and intellectual support of the Institute for Research on Exoplanets and the McGill Space Institute.}

	%%REFERENCES  
	
	%\clearpage

	\bibliographystyle{yahapj}
	\bibliography{1195}

\begin{thebibliography}{}
\providecommand\natexlab[1]{#1}
\providecommand\JournalTitle[1]{#1}

\bibitem[{{Batista} {et~al.}(2014){Batista}, {Beaulieu}, {Gould}, {Bennett},
  {Yee}, {Fukui}, {Gaudi}, {Sumi}, \& {Udalski}}]{Batista2014}
{Batista}, V., {Beaulieu}, J.-P., {Gould}, A., {et~al.} 2014,
  \JournalTitle{ApJ}, 780

\bibitem[{Beaulieu(2018)}]{Beaulieu2018}
Beaulieu, J.-P. 2018,
  \href{http://www.mdpi.com/2218-1997/4/4/61}{\JournalTitle{Universe}, 4}

\bibitem[{{Beaulieu} {et~al.}(2006){Beaulieu}, {Bennett}, {Fouqu{\'e}},
  {Williams}, {Dominik}, {J{\o}rgensen}, {Kubas}, {Cassan}, {Coutures},
  {Greenhill}, {Hill}, {Menzies}, {Sackett}, {Albrow}, {Brillant}, {Caldwell},
  {Calitz}, {Cook}, {Corrales}, {Desort}, {Dieters}, {Dominis}, {Donatowicz},
  {Hoffman}, {Kane}, {Marquette}, {Martin}, {Meintjes}, {Pollard}, {Sahu},
  {Vinter}, {Wambsganss}, {Woller}, {Horne}, {Steele}, {Bramich}, {Burgdorf},
  {Snodgrass}, {Bode}, {Udalski}, {Szyma{\'n}ski}, {Kubiak},
  {Wi{\c{e}}ckowski}, {Pietrzy{\'n}ski}, {Soszy{\'n}ski}, {Szewczyk},
  {Wyrzykowski}, {Paczy{\'n}ski}, {Abe}, {Bond}, {Britton}, {Gilmore},
  {Hearnshaw}, {Itow}, {Kamiya}, {Kilmartin}, {Korpela}, {Masuda}, {Matsubara},
  {Motomura}, {Muraki}, {Nakamura}, {Okada}, {Ohnishi}, {Rattenbury}, {Sako},
  {Sato}, {Sasaki}, {Sekiguchi}, {Sullivan}, {Tristram}, {Yock}, \&
  {Yoshioka}}]{Beaulieu2006}
{Beaulieu}, J.~P., {Bennett}, D.~P., {Fouqu{\'e}}, P., {et~al.} 2006,
  \href{http://dx.doi.org/10.1038/nature04441}{\JournalTitle{\nat}, 439, 437}

\bibitem[{{Beaulieu} {et~al.}(2016){Beaulieu}, {Bennett}, {Batista}, {Fukui},
  {Marquette}, {Brillant}, {Cole}, {Rogers}, {Sumi}, {Abe}, {Bhattacharya},
  {Koshimoto}, {Suzuki}, {Tristram}, {Han}, {Gould}, {Pogge}, \&
  {Yee}}]{Beaulieu2016}
{Beaulieu}, J.~P., {Bennett}, D.~P., {Batista}, V., {et~al.} 2016,
  \href{http://dx.doi.org/10.3847/0004-637X/824/2/83}{\JournalTitle{\apj}, 824,
  83}

\bibitem[{{Bennett}(2010)}]{Bennett2010model}
{Bennett}, D.~P. 2010,
  \href{http://dx.doi.org/10.1088/0004-637X/716/2/1408}{\JournalTitle{\apj},
  716, 1408}

\bibitem[{Bennett \& Rhie(1996)}]{Bennett1996}
Bennett, D.~P., \& Rhie, S.~H. 1996,
  \href{http://dx.doi.org/10.1086/178096}{\JournalTitle{ApJ}, 472, 660}

\bibitem[{{Bennett} {et~al.}(2008){Bennett}, {Bond}, {Udalski}, {Sumi}, {Abe},
  {Fukui}, {Furusawa}, {Hearnshaw}, {Holderness}, {Itow}, {Kamiya}, {Korpela},
  {Kilmartin}, {Lin}, {Ling}, {Masuda}, {Matsubara}, {Miyake}, {Muraki},
  {Nagaya}, {Okumura}, {Ohnishi}, {Perrott}, {Rattenbury}, {Sako}, {Saito},
  {Sato}, {Skuljan}, {Sullivan}, {Sweatman}, {Tristram}, {Yock}, {Kubiak},
  {Szyma{\'n}ski}, {Pietrzy{\'n}ski}, {Soszy{\'n}ski}, {Szewczyk},
  {Wyrzykowski}, {Ulaczyk}, {Batista}, {Beaulieu}, {Brillant}, {Cassan},
  {Fouqu{\'e}}, {Kervella}, {Kubas}, \& {Marquette}}]{Bennett2008}
{Bennett}, D.~P., {Bond}, I.~A., {Udalski}, A., {et~al.} 2008,
  \href{http://dx.doi.org/10.1086/589940}{\JournalTitle{\apj}, 684, 663}

\bibitem[{{Bennett} {et~al.}(2010){Bennett}, {Rhie}, {Nikolaev}, {Gaudi},
  {Udalski}, {Gould}, {Christie}, {Maoz}, {Dong}, {McCormick}, {Szyma{\'n}ski},
  {Tristram}, {Macintosh}, {Cook}, {Kubiak}, {Pietrzy{\'n}ski},
  {Soszy{\'n}ski}, {Szewczyk}, {Ulaczyk}, {Wyrzykowski}, {OGLE Collaboration},
  {DePoy}, {Han}, {Kaspi}, {Lee}, {Mallia}, {Natusch}, {Park}, {Pogge},
  {Polishook}, {{\ensuremath{\mu}}FUN Collaboration}, {Abe}, {Bond}, {Botzler},
  {Fukui}, {Hearnshaw}, {Itow}, {Kamiya}, {Korpela}, {Kilmartin}, {Lin},
  {Ling}, {Masuda}, {Matsubara}, {Motomura}, {Muraki}, {Nakamura}, {Okumura},
  {Ohnishi}, {Perrott}, {Rattenbury}, {Sako}, {Saito}, {Sato}, {Skuljan},
  {Sullivan}, {Sumi}, {Sweatman}, {Yock}, {MOA Collaboration}, {Albrow},
  {Allan}, {Beaulieu}, {Bramich}, {Burgdorf}, {Coutures}, {Dominik}, {Dieters},
  {Fouqu{\'e}}, {Greenhill}, {Horne}, {Snodgrass}, {Steele}, {Tsapras},
  {PLANET}, {RoboNet Collaborations}, {Chaboyer}, {Crocker}, \&
  {Frank}}]{Bennett2010apr}
{Bennett}, D.~P., {Rhie}, S.~H., {Nikolaev}, S., {et~al.} 2010,
  \href{http://dx.doi.org/10.1088/0004-637X/713/2/837}{\JournalTitle{\apj},
  713, 837}

\bibitem[{{Bennett} {et~al.}(2014){Bennett}, {Batista}, {Bond}, {Bennett},
  {Suzuki}, {Beaulieu}, {Udalski}, {Donatowicz}, {Bozza}, {Abe}, {Botzler},
  {Freeman}, {Fukunaga}, {Fukui}, {Itow}, {Koshimoto}, {Ling}, {Masuda},
  {Matsubara}, {Muraki}, {Namba}, {Ohnishi}, {Rattenbury}, {Saito}, {Sullivan},
  {Sumi}, {Sweatman}, {Tristram}, {Tsurumi}, {Wada}, {Yock}, {MOA
  Collaboration}, {Albrow}, {Bachelet}, {Brillant}, {Caldwell}, {Cassan},
  {Cole}, {Corrales}, {Coutures}, {Dieters}, {Dominis Prester}, {Fouqu{\'e}},
  {Greenhill}, {Horne}, {Koo}, {Kubas}, {Marquette}, {Martin}, {Menzies},
  {Sahu}, {Wambsganss}, {Williams}, {Zub}, {PLANET Collaboration}, {Choi},
  {DePoy}, {Dong}, {Gaudi}, {Gould}, {Han}, {Henderson}, {McGregor}, {Lee},
  {Pogge}, {Shin}, {Yee}, {{\ensuremath{\mu}}FUN Collaboration},
  {Szyma{\'n}ski}, {Skowron}, {Poleski}, {Koz{\l}owski}, {Wyrzykowski},
  {Kubiak}, {Pietrukowicz}, {Pietrzy{\'n}ski}, {Soszy{\'n}ski}, {Ulaczyk},
  {OGLE Collaboration}, {Tsapras}, {Street}, {Dominik}, {Bramich}, {Browne},
  {Hundertmark}, {Kains}, {Snodgrass}, {Steele}, {RoboNet Collaboration},
  {Dekany}, {Gonzalez}, {Heyrovsk{\'y}}, {Kandori}, {Kerins}, {Lucas},
  {Minniti}, {Nagayama}, {Rejkuba}, {Robin}, \& {Saito}}]{Bennett2014}
{Bennett}, D.~P., {Batista}, V., {Bond}, I.~A., {et~al.} 2014,
  \href{http://dx.doi.org/10.1088/0004-637X/785/2/155}{\JournalTitle{\apj},
  785, 155}

\bibitem[{{Bennett} {et~al.}(2015){Bennett}, {Bhattacharya}, {Anderson},
  {Bond}, {Anderson}, {Barry}, {Batista}, {Beaulieu}, {DePoy}, {Dong}, {Gaudi},
  {Gilbert}, {Gould}, {Pfeifle}, {Pogge}, {Suzuki}, {Terry}, \&
  {Udalski}}]{Bennett2015}
{Bennett}, D.~P., {Bhattacharya}, A., {Anderson}, J., {et~al.} 2015,
  \href{http://dx.doi.org/10.1088/0004-637X/808/2/169}{\JournalTitle{ApJ}, 808,
  169}

\bibitem[{Bennett {et~al.}(2016)Bennett, Rhie, Udalski, Gould, Tsapras, Kubas,
  Bond, Greenhill, Cassan, Rattenbury, Boyajian, Luhn, Penny, Anderson, Abe,
  Bhattacharya, Botzler, Donachie, Freeman, Fukui, Hirao, Itow, Koshimoto, Li,
  Ling, Masuda, Matsubara, Muraki, Nagakane, Ohnishi, Oyokawa, Perrott, Saito,
  Sharan, Sullivan, Sumi, Suzuki, Tristram, Yonehara, Yock, Collaboration),
  Szymański, Soszyński, Ulaczyk, Wyrzykowski, Collaboration), Allen, DePoy,
  Gal-Yam, Gaudi, Han, Monard, Ofek, Pogge, Collaboration), Street, Bramich,
  Dominik, Horne, Snodgrass, Steele, Collaboration), Albrow, Bachelet, Batista,
  Beaulieu, Brillant, Caldwell, Cole, Coutures, Dieters, Prester, Donatowicz,
  Fouqué, Hundertmark, Jørgensen, Kains, Kane, Marquette, Menzies, Pollard,
  Ranc, Sahu, Wambsganss, Williams, Zub, \& Collaboration)}]{Bennett_2016}
Bennett, D.~P., Rhie, S.~H., Udalski, A., {et~al.} 2016,
  \href{http://dx.doi.org/10.3847/0004-6256/152/5/125}{\JournalTitle{AJ}, 152,
  125}

\bibitem[{{Bennett} {et~al.}(2018{\natexlab{a}}){Bennett}, {Udalski}, {Bond},
  {Suzuki}, {Ryu}, {Abe}, {Barry}, {Bhattacharya}, {Donachie}, {Fukui},
  {Hirao}, {Kawasaki}, {Kondo}, {Koshimoto}, {Li}, {Matsubara}, {Miyazaki},
  {Muraki}, {Nagakane}, {Ohnishi}, {Ranc}, {Rattenbury}, {Suematsu}, {Sumi},
  {Tristram}, {Yonehara}, {MOA Collaboration}, {Szyma{\'n}ski},
  {Soszy{\'n}ski}, {Wyrzykowski}, {Ulaczyk}, {Poleski}, {Koz{\l}owski},
  {Pietrukowicz}, {Skowron}, {OGLE Collaboration}, {Shvartzvald}, {Maoz},
  {Kaspi}, {Friedmann}, {Wise Group}, {Batista}, {DePoy}, {Dong}, {Gaudi},
  {Gould}, {Han}, {Pogge}, {Tan}, {Yee}, \& {{\ensuremath{\mu}}FUN
  Collaboration}}]{Bennett2018}
{Bennett}, D.~P., {Udalski}, A., {Bond}, I.~A., {et~al.} 2018{\natexlab{a}},
  \href{http://dx.doi.org/10.3847/1538-3881/aad59c}{\JournalTitle{AJ}, 156,
  113}

\bibitem[{{Bennett} {et~al.}(2018{\natexlab{b}}){Bennett}, {Udalski}, {Han},
  {Bond}, {Beaulieu}, {Skowron}, {Gaudi}, {Koshimoto}, {Abe}, {Asakura},
  {Barry}, {Bhattacharya}, {Donachie}, {Evans}, {Fukui}, {Hirao}, {Itow}, {Li},
  {Ling}, {Masuda}, {Matsubara}, {Muraki}, {Nagakane}, {Ohnishi}, {Oyokawa},
  {Ranc}, {Rattenbury}, {Rosenthal}, {Saito}, {Sharan}, {Sullivan}, {Sumi},
  {Suzuki}, {Tristram}, {Yonehara}, {MOA Collaboration}, {Szyma{\'n}ski},
  {Poleski}, {Soszy{\'n}ski}, {Ulaczyk}, {Wyrzykowski}, {OGLE Collaboration},
  {DePoy}, {Gould}, {Pogge}, {Yee}, {{\ensuremath{\mu}}FUN Collaboration},
  {Albrow}, {Bachelet}, {Batista}, {Bowens-Rubin}, {Brillant}, {Caldwell},
  {Cole}, {Coutures}, {Dieters}, {Dominis Prester}, {Donatowicz}, {Fouqu{\'e}},
  {Horne}, {Hundertmark}, {Kains}, {Kane}, {Marquette}, {Menzies}, {Pollard},
  {Ranc}, {Sahu}, {Wambsganss}, {Williams}, {Zub}, \& {PLANET
  Collaboration}}]{Bennett2018b}
{Bennett}, D.~P., {Udalski}, A., {Han}, C., {et~al.} 2018{\natexlab{b}},
  \href{http://dx.doi.org/10.3847/1538-3881/aaadfa}{\JournalTitle{AJ}, 155,
  141}

\bibitem[{{Bennett} {et~al.}(2020){Bennett}, {Bhattacharya}, {Beaulieu},
  {Blackman}, {Vandorou}, {Terry}, {Cole}, {Henderson}, {Koshimoto}, {Lu},
  {Baptiste Marquette}, {Ranc}, \& {Udalski}}]{Bennett2020}
{Bennett}, D.~P., {Bhattacharya}, A., {Beaulieu}, J.-P., {et~al.} 2020,
  \href{http://dx.doi.org/10.3847/1538-3881/ab6212}{\JournalTitle{\aj}, 159,
  68}

\bibitem[{{Bennett} {et~al.}(2024){Bennett}, {Bhattacharya}, {Beaulieu},
  {Koshimoto}, {Blackman}, {Bond}, {Ranc}, {Rektsini}, {Terry}, {Vandorou},
  {Lu}, {Marquette}, {Olmschenk}, \& {Suzuki}}]{Bennett2024}
---. 2024, \href{http://dx.doi.org/10.3847/1538-3881/ad4880}{\JournalTitle{AJ},
  168, 15}

\bibitem[{{Bertin}(2010)}]{Bertin2010}
{Bertin}, E. 2010, {SWarp: Resampling and Co-adding FITS Images Together},
  Astrophysics Source Code Library,
  \href{http://arxiv.org/abs/1010.068}{{\sffamily ascl:1010.068}}

\bibitem[{{Bertin} \& {Arnouts}(1996)}]{Bertin1996}
{Bertin}, E., \& {Arnouts}, S. 1996,
  \href{http://dx.doi.org/10.1051/aas:1996164}{\JournalTitle{\aaps}, 117, 393}

\bibitem[{{Bessell} \& {Brett}(1988)}]{Bessell1988}
{Bessell}, M.~S., \& {Brett}, J.~M. 1988,
  \href{http://dx.doi.org/10.1086/132281}{\JournalTitle{\pasp}, 100, 1134}

\bibitem[{{Bhattacharya} {et~al.}(2018){Bhattacharya}, {Beaulieu}, {Bennett},
  {Anderson}, {Koshimoto}, {Lu}, {Batista}, {Blackman}, {Bond}, {Fukui},
  {Henderson}, {Hirao}, {Marquette}, {Mroz}, {Ranc}, \&
  {Udalski}}]{Bhattacharya2018}
{Bhattacharya}, A., {Beaulieu}, J.~P., {Bennett}, D.~P., {et~al.} 2018,
  \href{http://dx.doi.org/10.3847/1538-3881/aaed46}{\JournalTitle{\aj}, 156,
  289}

\bibitem[{{Blackman} {et~al.}(2021){Blackman}, {Beaulieu}, {Bennett},
  {Danielski}, {Alard}, {Cole}, {Vandorou}, {Ranc}, {Terry}, {Bhattacharya},
  {Bond}, {Bachelet}, {Veras}, {Koshimoto}, {Batista}, \&
  {Marquette}}]{Blackman2021}
{Blackman}, J.~W., {Beaulieu}, J.~P., {Bennett}, D.~P., {et~al.} 2021,
  \href{http://dx.doi.org/10.1038/s41586-021-03869-6}{\JournalTitle{\nat}, 598,
  272}

\bibitem[{{Bond} {et~al.}(2001){Bond}, {Abe}, {Dodd}, {Hearnshaw}, {Honda},
  {Jugaku}, {Kilmartin}, {Marles}, {Masuda}, {Matsubara}, {Muraki}, {Nakamura},
  {Nankivell}, {Noda}, {Noguchi}, {Ohnishi}, {Rattenbury}, {Reid}, {Saito},
  {Sato}, {Sekiguchi}, {Skuljan}, {Sullivan}, {Sumi}, {Takeuti}, {Watase},
  {Wilkinson}, {Yamada}, {Yanagisawa}, \& {Yock}}]{Bond2001}
{Bond}, I.~A., {Abe}, F., {Dodd}, R.~J., {et~al.} 2001,
  \href{http://dx.doi.org/10.1046/j.1365-8711.2001.04776.x}{\JournalTitle{\mnras},
  327, 868}

\bibitem[{{Bond} {et~al.}(2017){Bond}, {Bennett}, {Sumi}, {Udalski}, {Suzuki},
  {Rattenbury}, {Bozza}, {Koshimoto}, {Abe}, {Asakura}, {Barry},
  {Bhattacharya}, {Donachie}, {Evans}, {Fukui}, {Hirao}, {Itow}, {Li}, {Ling},
  {Masuda}, {Matsubara}, {Muraki}, {Nagakane}, {Ohnishi}, {Ranc}, {Saito},
  {Sharan}, {Sullivan}, {Tristram}, {Yamada}, {Yamada}, {Yonehara}, {Skowron},
  {Szyma{\'n}ski}, {Poleski}, {Mr{\'o}z}, {Soszy{\'n}ski}, {Pietrukowicz},
  {Koz{\l}owski}, {Ulaczyk}, \& {Pawlak}}]{Bond2017}
{Bond}, I.~A., {Bennett}, D.~P., {Sumi}, T., {et~al.} 2017,
  \href{http://dx.doi.org/10.1093/mnras/stx1049}{\JournalTitle{\mnras}, 469,
  2434}

\bibitem[{Boyajian {et~al.}(2014)Boyajian, van Belle, \& von
  Braun}]{Boyajian_2014}
Boyajian, T.~S., van Belle, G., \& von Braun, K. 2014,
  \href{http://dx.doi.org/10.1088/0004-6256/147/3/47}{\JournalTitle{ApJ}, 147,
  47}

\bibitem[{Burgasser {et~al.}(2006)Burgasser, Kirkpatrick, Cruz, Reid, Leggett,
  Liebert, Burrows, \& Brown}]{Burgasser_2006}
Burgasser, A.~J., Kirkpatrick, J.~D., Cruz, K.~L., {et~al.} 2006,
  \href{http://dx.doi.org/10.1086/506327}{\JournalTitle{AJS}, 166, 585}

\bibitem[{{Calchi Novati} {et~al.}(2015){Calchi Novati}, {Gould}, {Yee},
  {Beichman}, {Bryden}, {Carey}, {Fausnaugh}, {Gaudi}, {Henderson}, {Pogge},
  {Shvartzvald}, {Wibking}, {Zhu}, {Spitzer Team}, {Udalski}, {Poleski},
  {Pawlak}, {Szyma{\'n}ski}, {Skowron}, {Mr{\'o}z}, {Koz{\l}owski},
  {Wyrzykowski}, {Pietrukowicz}, {Pietrzy{\'n}ski}, {Soszy{\'n}ski}, {Ulaczyk},
  \& {OGLE Group}}]{CalchiNovati2015}
{Calchi Novati}, S., {Gould}, A., {Yee}, J.~C., {et~al.} 2015,
  \href{http://dx.doi.org/10.1088/0004-637X/814/2/92}{\JournalTitle{ApJ}, 814,
  92}

\bibitem[{Carmichael {et~al.}(2020)Carmichael, Quinn, Mustill, Huang, Zhou,
  Persson, Nielsen, Collins, Ziegler, Collins, Rodriguez, Shporer, Brahm, Mann,
  Bouchy, Fridlund, Stassun, Hellier, Seidel, Stalport, Udry, Pepe, Ireland,
  {\v{Z}}erjal, Brice{\~{n}}o, Law, Jord{\'{a}}n, Espinoza, Henning, Sarkis, \&
  Latham}]{Carmichael2020}
Carmichael, T.~W., Quinn, S.~N., Mustill, A.~J., {et~al.} 2020,
  \href{http://dx.doi.org/10.3847/1538-3881/ab9b84}{\JournalTitle{AJ}, 160, 53}

\bibitem[{Chin {et~al.}(2022)Chin, Cetre, Wizinowich, Ragland, Lilley,
  Wetherell, Surendran, Correia, Marin, Biasi, Pataunar, Pescoller, Glazebrook,
  Jameson, Gauvin, Rigaut, Gratadour, \& Bernard}]{Chin2022}
Chin, J. C.~Y., Cetre, S., Wizinowich, P., {et~al.}
  \href{http://dx.doi.org/10.1117/12.2629614}{2022, 12185, 121850V}

\bibitem[{{Chung} {et~al.}(2019){Chung}, {Gould}, {Skowron}, {Bond}, {Zhu},
  {Albrow}, {Jung}, {Han}, {Hwang}, {Ryu}, {Shin}, {Shvartzvald}, {Yee},
  {Zang}, {Cha}, {Kim}, {Kim}, {Kim}, {Kim}, {Lee}, {Lee}, {Lee}, {Park},
  {Pogge}, {KMTNet Collaboration}, {Udalski}, {Poleski}, {Mr{\'o}z},
  {Pietrukowicz}, {Szyma{\'n}ski}, {Soszy{\'n}ski}, {Koz{\l}owski}, {Ulaczyk},
  {Pawlak}, {OGLE Collaboration}, {Beichman}, {Bryden}, {Calchi Novati},
  {Carey}, {Gaudi}, {Henderson}, {Spitzer Team}, {Abe}, {Barry}, {Bennett},
  {Bhattacharya}, {Donachie}, {Fukui}, {Hirao}, {Itow}, {Kawasaki}, {Kondo},
  {Koshimoto}, {Li}, {Matsubara}, {Muraki}, {Miyazaki}, {Nagakane}, {Ranc},
  {Rattenbury}, {Suematsu}, {Sullivan}, {Sumi}, {Suzuki}, {Tristram},
  {Yonehara}, \& {MOA colllaboration}}]{Chun2019}
{Chung}, S.-J., {Gould}, A., {Skowron}, J., {et~al.} 2019,
  \href{http://dx.doi.org/10.3847/1538-4357/aaf861}{\JournalTitle{\apj}, 871,
  179}

\bibitem[{{Cumming} {et~al.}(2008){Cumming}, {Butler}, {Marcy}, {Vogt},
  {Wright}, \& {Fischer}}]{Cumming2008}
{Cumming}, A., {Butler}, R.~P., {Marcy}, G.~W., {et~al.} 2008,
  \href{http://dx.doi.org/10.1086/588487}{\JournalTitle{Publications of the
  Astronomical Society of the Pacific}, 120, 531}

\bibitem[{Dang {et~al.}(2020)Dang, Calchi Novati, Carey, \& Cowan}]{Dang2020}
Dang, L., Calchi Novati, S., Carey, S., \& Cowan, N.~B. 2020,
  \href{http://dx.doi.org/10.1093/mnras/staa2245}{\JournalTitle{MNRAS}, 497,
  5309}

\bibitem[{{Delfosse} {et~al.}(2000){Delfosse}, {Forveille}, {S{\'e}gransan},
  {Beuzit}, {Udry}, {Perrier}, \& {Mayor}}]{Delfosse2000}
{Delfosse}, X., {Forveille}, T., {S{\'e}gransan}, D., {et~al.} 2000,
  \JournalTitle{AAP}, 364, 217

\bibitem[{{Dominik}(1998)}]{Dominik1998}
{Dominik}, M. 1998,
  \href{http://dx.doi.org/10.48550/arXiv.astro-ph/9702039}{\JournalTitle{\aap},
  329, 361}

\bibitem[{Drimmel \& Spergel(2001)}]{Drimmel2001}
Drimmel, R., \& Spergel, D.~N. 2001,
  \href{http://dx.doi.org/10.1086/321556}{\JournalTitle{ApJ}, 556, 181}

\bibitem[{{Fazio} {et~al.}(2004){Fazio}, {Hora}, {Allen}, {Ashby}, {Barmby},
  {Deutsch}, {Huang}, {Kleiner}, {Marengo}, {Megeath}, {Melnick}, {Pahre},
  {Patten}, {Polizotti}, {Smith}, {Taylor}, {Wang}, {Willner}, {Hoffmann},
  {Pipher}, {Forrest}, {McMurty}, {McCreight}, {McKelvey}, {McMurray}, {Koch},
  {Moseley}, {Arendt}, {Mentzell}, {Marx}, {Losch}, {Mayman}, {Eichhorn},
  {Krebs}, {Jhabvala}, {Gezari}, {Fixsen}, {Flores}, {Shakoorzadeh}, {Jungo},
  {Hakun}, {Workman}, {Karpati}, {Kichak}, {Whitley}, {Mann}, {Tollestrup},
  {Eisenhardt}, {Stern}, {Gorjian}, {Bhattacharya}, {Carey}, {Nelson},
  {Glaccum}, {Lacy}, {Lowrance}, {Laine}, {Reach}, {Stauffer}, {Surace},
  {Wilson}, {Wright}, {Hoffman}, {Domingo}, \& {Cohen}}]{Fazio2004}
{Fazio}, G.~G., {Hora}, J.~L., {Allen}, L.~E., {et~al.} 2004,
  \href{http://dx.doi.org/10.1086/422843}{\JournalTitle{ApJS}, 154, 10}

\bibitem[{{Furusawa} {et~al.}(2013){Furusawa}, {Udalski}, {Sumi}, {Bennett},
  {Bond}, {Gould}, {J{\o}rgensen}, {Snodgrass}, {Dominis Prester}, {Albrow},
  {Abe}, {Botzler}, {Chote}, {Freeman}, {Fukui}, {Harris}, {Itow}, {Ling},
  {Masuda}, {Matsubara}, {Miyake}, {Muraki}, {Ohnishi}, {Rattenbury}, {Saito},
  {Sullivan}, {Suzuki}, {Sweatman}, {Tristram}, {Wada}, {Yock}, {MOA
  Collaboration}, {Szyma{\'n}ski}, {Soszy{\'n}ski}, {Kubiak}, {Poleski},
  {Ulaczyk}, {Pietrzy{\'n}ski}, {Wyrzykowski}, {OGLE Collaboration}, {Choi},
  {Christie}, {DePoy}, {Dong}, {Drummond}, {Gaudi}, {Han}, {Hung}, {Hwang},
  {Lee}, {McCormick}, {Moorhouse}, {Natusch}, {Nola}, {Ofek}, {Pogge}, {Shin},
  {Skowron}, {Thornley}, {Yee}, {{\ensuremath{\mu}}FUN Collaboration},
  {Alsubai}, {Bozza}, {Browne}, {Burgdorf}, {Calchi Novati}, {Dodds},
  {Dominik}, {Finet}, {Gerner}, {Hardis}, {Harps{\o}e}, {Hinse}, {Hundertmark},
  {Kains}, {Kerins}, {Liebig}, {Mancini}, {Mathiasen}, {Penny}, {Proft},
  {Rahvar}, {Ricci}, {Scarpetta}, {Sch{\"a}fer}, {Sch{\"o}nebeck},
  {Southworth}, {Surdej}, {Wambsganss}, {MiNDSTEp Consortium}, {Street},
  {Bramich}, {Steele}, {Tsapras}, {RoboNet Collaboration}, {Horne},
  {Donatowicz}, {Sahu}, {Bachelet}, {Batista}, {Beatty}, {Beaulieu}, {Bennett},
  {Black}, {Bowens-Rubin}, {Brillant}, {Caldwell}, {Cassan}, {Cole},
  {Corrales}, {Coutures}, {Dieters}, {Fouqu{\'e}}, {Greenhill}, {Henderson},
  {Kubas}, {Marquette}, {Martin}, {Menzies}, {Shappee}, {Williams}, {Wouters},
  {van Saders}, {Zellem}, {Zub}, \& {PLANET Collaboration}}]{Furusawa2013}
{Furusawa}, K., {Udalski}, A., {Sumi}, T., {et~al.} 2013,
  \href{http://dx.doi.org/10.1088/0004-637X/779/2/91}{\JournalTitle{ApJ}, 779,
  91}

\bibitem[{Gaudi(1998)}]{Gaudi1998}
Gaudi, B.~S. 1998, \href{http://dx.doi.org/10.1086/306256}{\JournalTitle{ApJ},
  506, 533}

\bibitem[{{Gaudi} {et~al.}(2008){Gaudi}, {Bennett}, {Udalski}, {Gould},
  {Christie}, {Maoz}, {Dong}, {McCormick}, {Szyma{\'n}ski}, {Tristram},
  {Nikolaev}, {Paczy{\'n}ski}, {Kubiak}, {Pietrzy{\'n}ski}, {Soszy{\'n}ski},
  {Szewczyk}, {Ulaczyk}, {Wyrzykowski}, {OGLE Collaboration}, {DePoy}, {Han},
  {Kaspi}, {Lee}, {Mallia}, {Natusch}, {Pogge}, {Park}, {{\ensuremath{\mu}}-Fun
  Collabortion}, {Abe}, {Bond}, {Botzler}, {Fukui}, {Hearnshaw}, {Itow},
  {Kamiya}, {Korpela}, {Kilmartin}, {Lin}, {Masuda}, {Matsubara}, {Motomura},
  {Muraki}, {Nakamura}, {Okumura}, {Ohnishi}, {Rattenbury}, {Sako}, {Saito},
  {Sato}, {Skuljan}, {Sullivan}, {Sumi}, {Sweatman}, {Yock}, {MOA
  Collaboration}, {Albrow}, {Allan}, {Beaulieu}, {Burgdorf}, {Cook},
  {Coutures}, {Dominik}, {Dieters}, {Fouqu{\'e}}, {Greenhill}, {Horne},
  {Steele}, {Tsapras}, {Planet Collaboration}, {RoboNet Collaborations},
  {Chaboyer}, {Crocker}, {Frank}, \& {Macintosh}}]{Gaudi2008}
{Gaudi}, B.~S., {Bennett}, D.~P., {Udalski}, A., {et~al.} 2008,
  \href{http://dx.doi.org/10.1126/science.1151947}{\JournalTitle{Science}, 319,
  927}

\bibitem[{{Gould}(1992)}]{Gould1992A}
{Gould}, A. 1992, \href{http://dx.doi.org/10.1086/171443}{\JournalTitle{\apj},
  392, 442}

\bibitem[{{Gould} \& {Loeb}(1992)}]{Gould1992}
{Gould}, A., \& {Loeb}, A. 1992,
  \href{http://dx.doi.org/10.1086/171700}{\JournalTitle{\apj}, 396, 104}

\bibitem[{{Gould} {et~al.}(2023){Gould}, {Shvartzvald}, {Zhang}, {Yee}, {Calchi
  Novati}, {Zang}, \& {Ofek}}]{Gould2023}
{Gould}, A., {Shvartzvald}, Y., {Zhang}, J., {et~al.} 2023,
  \href{http://dx.doi.org/10.3847/1538-3881/aced3c}{\JournalTitle{AJ}, 166,
  145}

\bibitem[{{Gould} {et~al.}(2010){Gould}, {Dong}, {Gaudi}, {Udalski}, {Bond},
  {Greenhill}, {Street}, {Dominik}, {Sumi}, {Szyma{\'n}ski}, {Han}, {Allen},
  {Bolt}, {Bos}, {Christie}, {DePoy}, {Drummond}, {Eastman}, {Gal-Yam},
  {Higgins}, {Janczak}, {Kaspi}, {Koz{\l}owski}, {Lee}, {Mallia}, {Maury},
  {Maoz}, {McCormick}, {Monard}, {Moorhouse}, {Morgan}, {Natusch}, {Ofek},
  {Park}, {Pogge}, {Polishook}, {Santallo}, {Shporer}, {Spector}, {Thornley},
  {Yee}, {{$\mu$}FUN Collaboration}, {Kubiak}, {Pietrzy{\'n}ski},
  {Soszy{\'n}ski}, {Szewczyk}, {Wyrzykowski}, {Ulaczyk}, {Poleski}, {OGLE
  Collaboration}, {Abe}, {Bennett}, {Botzler}, {Douchin}, {Freeman}, {Fukui},
  {Furusawa}, {Hearnshaw}, {Hosaka}, {Itow}, {Kamiya}, {Kilmartin}, {Korpela},
  {Lin}, {Ling}, {Makita}, {Masuda}, {Matsubara}, {Miyake}, {Muraki}, {Nagaya},
  {Nishimoto}, {Ohnishi}, {Okumura}, {Perrott}, {Philpott}, {Rattenbury},
  {Saito}, {Sako}, {Sullivan}, {Sweatman}, {Tristram}, {von Seggern}, {Yock},
  {MOA Collaboration}, {Albrow}, {Batista}, {Beaulieu}, {Brillant}, {Caldwell},
  {Calitz}, {Cassan}, {Cole}, {Cook}, {Coutures}, {Dieters}, {Dominis Prester},
  {Donatowicz}, {Fouqu{\'e}}, {Hill}, {Hoffman}, {Jablonski}, {Kane}, {Kains},
  {Kubas}, {Marquette}, {Martin}, {Martioli}, {Meintjes}, {Menzies},
  {Pedretti}, {Pollard}, {Sahu}, {Vinter}, {Wambsganss}, {Watson}, {Williams},
  {Zub}, {PLANET Collaboration}, {Allan}, {Bode}, {Bramich}, {Burgdorf},
  {Clay}, {Fraser}, {Hawkins}, {Horne}, {Kerins}, {Lister}, {Mottram},
  {Saunders}, {Snodgrass}, {Steele}, {Tsapras}, {RoboNet Collaboration},
  {J{\o}rgensen}, {Anguita}, {Bozza}, {Calchi Novati}, {Harps{\o}e}, {Hinse},
  {Hundertmark}, {Kj{\ae}rgaard}, {Liebig}, {Mancini}, {Masi}, {Mathiasen},
  {Rahvar}, {Ricci}, {Scarpetta}, {Southworth}, {Surdej}, {Th{\"o}ne}, \&
  {MiNDSTEp Consortium}}]{Gould2010}
{Gould}, A., {Dong}, S., {Gaudi}, B.~S., {et~al.} 2010, \JournalTitle{ApJ},
  720, 1073

\bibitem[{{Griest} \& {Hu}(1992)}]{Griest1992}
{Griest}, K., \& {Hu}, W. 1992,
  \href{http://dx.doi.org/10.1086/171793}{\JournalTitle{\apj}, 397, 362}

\bibitem[{{Han} \& {Gould}(1995)}]{Han1995}
{Han}, C., \& {Gould}, A. 1995,
  \href{http://dx.doi.org/10.1086/176076}{\JournalTitle{\apj}, 449, 521}

\bibitem[{{Han} {et~al.}(2023{\natexlab{a}}){Han}, {Zang}, {Jung}, {Bond},
  {Chung}, {Albrow}, {Gould}, {Hwang}, {Ryu}, {Shin}, {Shvartzvald}, {Yang},
  {Yee}, {Cha}, {Kim}, {Kim}, {Kim}, {Lee}, {Lee}, {Lee}, {Park}, {Pogge},
  {Monard}, {Qian}, {Liu}, {Maoz}, {Penny}, {Zhu}, {Abe}, {Barry}, {Bennett},
  {Bhattacharya}, {Fujii}, {Fukui}, {Hamada}, {Hirao}, {Ishitani Silva},
  {Itow}, {Kirikawa}, {Kondo}, {Koshimoto}, {Matsubara}, {Miyazaki}, {Muraki},
  {Olmschenk}, {Ranc}, {Rattenbury}, {Satoh}, {Sumi}, {Suzuki}, {Tomoyoshi},
  {Tristram}, {Vandorou}, {Yama}, \& {Yamashita}}]{Han2023}
{Han}, C., {Zang}, W., {Jung}, Y.~K., {et~al.} 2023{\natexlab{a}},
  \href{http://dx.doi.org/10.1051/0004-6361/202347366}{\JournalTitle{\aap},
  678, A101}

\bibitem[{{Han} {et~al.}(2023{\natexlab{b}}){Han}, {Udalski}, {Jung}, {Kim},
  {Yang}, {Albrow}, {Chung}, {Gould}, {Hwang}, {Kim}, {Lee}, {Ryu},
  {Shvartzvald}, {Shin}, {Yee}, {Zang}, {Cha}, {Kim}, {Kim}, {Lee}, {Lee},
  {Park}, {Pogge}, {Kim}, {Kim}, {Mr{\'o}z}, {Szyma{\'n}ski}, {Skowron},
  {Poleski}, {Soszy{\'n}ski}, {Pietrukowicz}, {Koz{\l}owski}, {Rybicki},
  {Iwanek}, {Ulaczyk}, {Wrona}, \& {Gromadzki}}]{Han2023a}
{Han}, C., {Udalski}, A., {Jung}, Y.~K., {et~al.} 2023{\natexlab{b}},
  \href{http://dx.doi.org/10.1051/0004-6361/202245525}{\JournalTitle{\aap},
  670, A172}

\bibitem[{Henry {et~al.}(1999)Henry, Franz, Wasserman, Benedict, Shelus, Ianna,
  Kirkpatrick, \& Donald W.~McCarthy}]{Henry1999}
Henry, T.~J., Franz, O.~G., Wasserman, L.~H., {et~al.} 1999,
  \href{http://dx.doi.org/10.1086/306793}{\JournalTitle{ApJ}, 512, 864}

\bibitem[{{Henry} \& {McCarthy}(1993)}]{Henry1993}
{Henry}, T.~J., \& {McCarthy}, Donald~W., J. 1993,
  \href{http://dx.doi.org/10.1086/116685}{\JournalTitle{AJ}, 106, 773}

\bibitem[{{Ida} \& {Lin}(2004)}]{Ida2004}
{Ida}, S., \& {Lin}, D.~N.~C. 2004,
  \href{http://dx.doi.org/10.1086/381724}{\JournalTitle{\apj}, 604, 388}

\bibitem[{{Kennedy} {et~al.}(2006){Kennedy}, {Kenyon}, \&
  {Bromley}}]{Kennedy2006}
{Kennedy}, G.~M., {Kenyon}, S.~J., \& {Bromley}, B.~C. 2006,
  \href{http://dx.doi.org/10.1086/508882}{\JournalTitle{\apj}, 650, L139}

\bibitem[{{Kim} {et~al.}(2016){Kim}, {Lee}, {Park}, {Kim}, {Cha}, {Lee}, {Han},
  {Chun}, \& {Yuk}}]{Kim2016}
{Kim}, S.-L., {Lee}, C.-U., {Park}, B.-G., {et~al.} 2016,
  \href{http://dx.doi.org/10.5303/JKAS.2016.49.1.37}{\JournalTitle{Journal of
  Korean Astronomical Society}, 49, 37}

\bibitem[{{Koshimoto} {et~al.}(2021){Koshimoto}, {Baba}, \&
  {Bennett}}]{Koshimoto2021a}
{Koshimoto}, N., {Baba}, J., \& {Bennett}, D.~P. 2021,
  \href{http://dx.doi.org/10.3847/1538-4357/ac07a8}{\JournalTitle{\apj}, 917,
  78}

\bibitem[{{Koshimoto} \& {Bennett}(2020)}]{Koshimoto2020}
{Koshimoto}, N., \& {Bennett}, D.~P. 2020,
  \href{http://dx.doi.org/10.3847/1538-3881/abaf4e}{\JournalTitle{AJ}, 160,
  177}

\bibitem[{{Kroupa}(2001)}]{Kroupa2001}
{Kroupa}, P. 2001,
  \href{http://dx.doi.org/10.1046/j.1365-8711.2001.04022.x}{\JournalTitle{MNRAS},
  322, 231}

\bibitem[{{Lecavelier des Etangs} \& {Lissauer}(2022)}]{IAU2022}
{Lecavelier des Etangs}, A., \& {Lissauer}, J.~J. 2022,
  \href{http://dx.doi.org/10.1016/j.newar.2022.101641}{\JournalTitle{NAR}, 94,
  101641}

\bibitem[{{Lissauer}(1993)}]{Lissauer1993}
{Lissauer}, J.~J. 1993,
  \href{http://dx.doi.org/10.1146/annurev.aa.31.090193.001021}{\JournalTitle{Annual
  Review of Astronomy and Astrophysics}, 31, 129}

\bibitem[{Lu {et~al.}(2021)Lu, Gautam, Chu, Terry, \& Do}]{Lu_code_2022}
Lu, J.~R., Gautam, A.~K., Chu, D., Terry, S.~K., \& Do, T. 2021,
  {Keck-DataReductionPipelines/KAI: v1.0.0 Release of KAI}

\bibitem[{{Mao} \& {Paczynski}(1991)}]{Mao1991}
{Mao}, S., \& {Paczynski}, B. 1991,
  \href{http://dx.doi.org/10.1086/186066}{\JournalTitle{\apjl}, 374, L37}

\bibitem[{{Minniti} {et~al.}(2010){Minniti}, {Lucas}, {Emerson}, {Saito},
  {Hempel}, {Pietrukowicz}, {Ahumada}, {Alonso}, {Alonso-Garcia}, {Arias},
  {Bandyopadhyay}, {Barb{\'a}}, {Barbuy}, {Bedin}, {Bica}, {Borissova},
  {Bronfman}, {Carraro}, {Catelan}, {Clari{\'a}}, {Cross}, {de Grijs},
  {D{\'e}k{\'a}ny}, {Drew}, {Fari{\~n}a}, {Feinstein}, {Fern{\'a}ndez
  Laj{\'u}s}, {Gamen}, {Geisler}, {Gieren}, {Goldman}, {Gonzalez}, {Gunthardt},
  {Gurovich}, {Hambly}, {Irwin}, {Ivanov}, {Jord{\'a}n}, {Kerins}, {Kinemuchi},
  {Kurtev}, {L{\'o}pez-Corredoira}, {Maccarone}, {Masetti}, {Merlo},
  {Messineo}, {Mirabel}, {Monaco}, {Morelli}, {Padilla}, {Palma}, {Parisi},
  {Pignata}, {Rejkuba}, {Roman-Lopes}, {Sale}, {Schreiber}, {Schr{\"o}der},
  {Smith}, {}, {Soto}, {Tamura}, {Tappert}, {Thompson}, {Toledo}, {Zoccali}, \&
  {Pietrzynski}}]{Minniti2010}
{Minniti}, D., {Lucas}, P.~W., {Emerson}, J.~P., {et~al.} 2010,
  \href{http://dx.doi.org/10.1016/j.newast.2009.12.002}{\JournalTitle{\na}, 15,
  433}

\bibitem[{{Mulders} {et~al.}(2015){Mulders}, {Pascucci}, \&
  {Apai}}]{Mulders2015}
{Mulders}, G.~D., {Pascucci}, I., \& {Apai}, D. 2015,
  \href{http://dx.doi.org/10.1088/0004-637X/798/2/112}{\JournalTitle{\apj},
  798, 112}

\bibitem[{{Muzic} {et~al.}(2017){Muzic}, {Schodel}, {Scholz}, {Geers},
  {Jayawardhana}, {Ascenso}, \& {Cieza}}]{Muzic}
{Muzic}, K., {Schodel}, R., {Scholz}, A., {et~al.} 2017,
  \href{http://dx.doi.org/10.1093/mnras/stx1906}{\JournalTitle{MNRAS}, 471,
  3699}

\bibitem[{{Nishiyama} {et~al.}(2006){Nishiyama}, {Nagata}, {Kusakabe},
  {Matsunaga}, {Naoi}, {Kato}, {Nagashima}, {Sugitani}, {Tamura}, {Tanab{\'e}},
  \& {Sato}}]{Nishiyama2006}
{Nishiyama}, S., {Nagata}, T., {Kusakabe}, N., {et~al.} 2006,
  \href{http://dx.doi.org/10.1086/499038}{\JournalTitle{ApJ}, 638, 839}

\bibitem[{{Pascucci} {et~al.}(2018){Pascucci}, {Mulders}, {Gould}, \&
  {Fernandes}}]{Pascucci2018}
{Pascucci}, I., {Mulders}, G.~D., {Gould}, A., \& {Fernandes}, R. 2018,
  \href{http://dx.doi.org/10.3847/2041-8213/aab6ac}{\JournalTitle{ApJL}, 856,
  L28}

\bibitem[{{Quenouille}(1949)}]{Quenouille1949}
{Quenouille}, M.~H. 1949,
  \href{http://dx.doi.org/10.1214/aoms/1177729989}{\JournalTitle{Ann. Math.
  Stat.}, 20, 355}

\bibitem[{{Quenouille}(1956)}]{Quenouille1956}
---. 1956,
  \href{http://dx.doi.org/10.1093/biomet/43.3-4.353}{\JournalTitle{Biometrika},
  43, 353}

\bibitem[{{Reid} {et~al.}(2002){Reid}, {Gizis}, \& {Hawley}}]{Reid2002}
{Reid}, I.~N., {Gizis}, J.~E., \& {Hawley}, S.~L. 2002,
  \href{http://dx.doi.org/10.1086/343777}{\JournalTitle{\aj}, 124, 2721}

\bibitem[{{Shvartzvald} {et~al.}(2017){Shvartzvald}, {Yee}, {Calchi Novati},
  {Gould}, {Lee}, {Beichman}, {Bryden}, {Carey}, {Gaudi}, {Henderson}, {Zhu},
  {Spitzer Team}, {Albrow}, {Cha}, {Chung}, {Han}, {Hwang}, {Jung}, {Kim},
  {Kim}, {Kim}, {Lee}, {Park}, {Pogge}, {Ryu}, {Shin}, \& {KMTNet
  Group}}]{Shvartzvald2017}
{Shvartzvald}, Y., {Yee}, J.~C., {Calchi Novati}, S., {et~al.} 2017,
  \href{http://dx.doi.org/10.3847/2041-8213/aa6d09}{\JournalTitle{\apj}, 840,
  L3}

\bibitem[{{Shvartzvald} {et~al.}(2019){Shvartzvald}, {Yee}, {Skowron}, {Lee},
  {Udalski}, {Calchi Novati}, {Bozza}, {Beichman}, {Bryden}, {Carey}, {Gaudi},
  {Henderson}, {Zhu}, {Spitzer Team}, {Bachelet}, {Bolt}, {Christie}, {Maoz},
  {Natusch}, {Pogge}, {Street}, {Tan}, {Tsapras}, {LCO}, {{\ensuremath{\mu}}FUN
  Follow-up Teams}, {Pietrukowicz}, {Soszy{\'n}ski}, {Szyma{\'n}ski},
  {Mr{\'o}z}, {Poleski}, {Koz{\l}owski}, {Ulaczyk}, {Pawlak}, {Rybicki},
  {Iwanek}, {OGLE Collaboration}, {Albrow}, {Cha}, {Chung}, {Gould}, {Han},
  {Hwang}, {Jung}, {Kim}, {Kim}, {Kim}, {Lee}, {Lee}, {Park}, {Ryu}, {Shin},
  {Zang}, {KMTNet Collaboration}, {Dominik}, {Helling}, {Hundertmark},
  {J{\o}rgensen}, {Longa-Pe{\~n}a}, {Lowry}, {Sajadian}, {Burgdorf},
  {Campbell-White}, {Ciceri}, {Evans}, {Fujii}, {Hinse}, {Rahvar}, {Rabus},
  {Skottfelt}, {Snodgrass}, {Southworth}, \& {MiNDSTEp
  Collaboration}}]{Shvartzvald2019}
{Shvartzvald}, Y., {Yee}, J.~C., {Skowron}, J., {et~al.} 2019,
  \href{http://dx.doi.org/10.3847/1538-3881/aafe12}{\JournalTitle{AJ}, 157,
  106}

\bibitem[{{Stanek} {et~al.}(1994){Stanek}, {Mateo}, {Udalski}, {Szymanski},
  {Kaluzny}, \& {Kubiak}}]{Stanek1994}
{Stanek}, K.~Z., {Mateo}, M., {Udalski}, A., {et~al.} 1994,
  \href{http://dx.doi.org/10.1086/187416}{\JournalTitle{\apjl}, 429, L73}

\bibitem[{{Stetson}(1987)}]{Stetson1987}
{Stetson}, P.~B. 1987,
  \href{http://dx.doi.org/10.1086/131977}{\JournalTitle{\pasp}, 99, 191}

\bibitem[{{Surot} {et~al.}(2019){Surot}, {Valenti}, {Hidalgo}, {Zoccali},
  {S{\"o}kmen}, {Rejkuba}, {Minniti}, {Gonzalez}, {Cassisi}, {Renzini}, \&
  {Weiss}}]{Surot2019}
{Surot}, F., {Valenti}, E., {Hidalgo}, S.~L., {et~al.} 2019,
  \href{http://dx.doi.org/10.1051/0004-6361/201833550}{\JournalTitle{AAP}, 623,
  A168}

\bibitem[{{Suzuki} {et~al.}(2016){Suzuki}, {Bennett}, {Sumi}, {Bond}, {Rogers},
  {Abe}, {Asakura}, {Bhattacharya}, {Donachie}, {Freeman}, {Fukui}, {Hirao},
  {Itow}, {Koshimoto}, {Li}, {Ling}, {Masuda}, {Matsubara}, {Muraki},
  {Nagakane}, {Onishi}, {Oyokawa}, {Rattenbury}, {Saito}, {Sharan}, {Shibai},
  {Sullivan}, {Tristram}, {Yonehara}, \& {MOA Collaboration}}]{Suzuki2016}
{Suzuki}, D., {Bennett}, D.~P., {Sumi}, T., {et~al.} 2016,
  \href{http://dx.doi.org/10.3847/1538-4357/833/2/145}{\JournalTitle{\apj},
  833, 145}

\bibitem[{{Suzuki} {et~al.}(2018){Suzuki}, {Bennett}, {Udalski}, {Bond},
  {Sumi}, {Han}, {Kim}, {Abe}, {Asakura}, {Barry}, {Bhattacharya}, {Donachie},
  {Freeman}, {Fukui}, {Hirao}, {Itow}, {Koshimoto}, {Li}, {Ling}, {Masuda},
  {Matsubara}, {Muraki}, {Nagakane}, {Onishi}, {Oyokawa}, {Ranc}, {Rattenbury},
  {Saito}, {Sharan}, {Sullivan}, {Tristram}, {Yonehara}, {MOA Collaboration},
  {Poleski}, {Mr{\'o}z}, {Skowron}, {Szyma{\'n}ski}, {Soszy{\'n}ski},
  {Koz{\l}owski}, {Pietrukowicz}, {Wyrzykowski}, {Ulaczyk}, \& {OGLE
  Collaboration}}]{Suzuki2018}
{Suzuki}, D., {Bennett}, D.~P., {Udalski}, A., {et~al.} 2018,
  \href{http://dx.doi.org/10.3847/1538-3881/aabd7a}{\JournalTitle{\aj}, 155,
  263}

\bibitem[{{Terry} {et~al.}(2020){Terry}, {Barry}, {Bennett}, {Bhattacharya},
  {Anderson}, \& {Penny}}]{Terry2020}
{Terry}, S.~K., {Barry}, R.~K., {Bennett}, D.~P., {et~al.} 2020,
  \href{http://dx.doi.org/10.3847/1538-4357/ab629b}{\JournalTitle{\apj}, 889,
  126}

\bibitem[{{Terry} {et~al.}(2021){Terry}, {Bhattacharya}, {Bennett}, {Beaulieu},
  {Koshimoto}, {Blackman}, {Bond}, {Cole}, {Henderson}, {Lu}, {Marquette},
  {Ranc}, \& {Vandorou}}]{Terrry2021}
{Terry}, S.~K., {Bhattacharya}, A., {Bennett}, D.~P., {et~al.} 2021,
  \href{http://dx.doi.org/10.3847/1538-3881/abcc60}{\JournalTitle{\aj}, 161,
  54}

\bibitem[{{Terry} {et~al.}(2022){Terry}, {Bennett}, {Bhattacharya},
  {Koshimoto}, {Beaulieu}, {Blackman}, {Bond}, {Cole}, {Lu}, {Marquette},
  {Ranc}, {Rektsini}, \& {Vandorou}}]{Terry2022}
{Terry}, S.~K., {Bennett}, D.~P., {Bhattacharya}, A., {et~al.} 2022,
  \href{http://dx.doi.org/10.3847/1538-3881/ac9518}{\JournalTitle{AJ}, 164,
  217}

\bibitem[{Tukey(1958)}]{Tukey1958}
Tukey, J.~W. 1958,
  \href{http://dx.doi.org/10.1214/aoms/1177706635}{\JournalTitle{The Annals of
  Mathematical Statistics}, 29, 581 }

\bibitem[{{Udalski}(2003)}]{Udalski2003}
{Udalski}, A. 2003, \JournalTitle{\actaa}, 53, 291

\bibitem[{{Udalski} {et~al.}(1994){Udalski}, {Szymanski}, {Kaluzny}, {Kubiak},
  {Mateo}, {Krzeminski}, \& {Paczynski}}]{Udalski1994}
{Udalski}, A., {Szymanski}, M., {Kaluzny}, J., {et~al.} 1994,
  \JournalTitle{\actaa}, 44, 227

\bibitem[{{Udalski} {et~al.}(2018){Udalski}, {Ryu}, {Sajadian}, {Gould},
  {Mr{\'o}z}, {Poleski}, {Szyma{\'n}ski}, {Skowron}, {Soszy{\'n}ski},
  {Koz{\l}owski}, {Pietrukowicz}, {Ulaczyk}, {Pawlak}, {Rybicki}, {Iwanek},
  {Albrow}, {Chung}, {Han}, {Hwang}, {Jung}, {Shin}, {Shvartzvald}, {Yee},
  {Zang}, {Zhu}, {Cha}, {Kim}, {Kim}, {Kim}, {Lee}, {Lee}, {Lee}, {Park},
  {Pogge}, {Bozza}, {Dominik}, {Helling}, {Hundertmark}, {J{\o}rgensen},
  {Longa-Pe{\~n}a}, {Lowry}, {Burgdorf}, {Campbell-White}, {Ciceri}, {Evans},
  {Figuera Jaimes}, {Fujii}, {Haikala}, {Henning}, {Hinse}, {Mancini},
  {Peixinho}, {Rahvar}, {Rabus}, {Skottfelt}, {Snodgrass}, {Southworth}, \&
  {von Essen}}]{Udalski2018}
{Udalski}, A., {Ryu}, Y.~H., {Sajadian}, S., {et~al.} 2018,
  \href{http://dx.doi.org/10.32023/0001-5237/68.1.1}{\JournalTitle{ACTAA}, 68,
  1}

\bibitem[{{Vandorou} {et~al.}(2020){Vandorou}, {Bennett}, {Beaulieu}, {Alard},
  {Blackman}, {Cole}, {Bhattacharya}, {Bond}, {Koshimoto}, \&
  {Marquette}}]{Vandorou2020}
{Vandorou}, A., {Bennett}, D.~P., {Beaulieu}, J.-P., {et~al.} 2020,
  \href{http://dx.doi.org/10.3847/1538-3881/aba2d3}{\JournalTitle{\aj}, 160,
  121}

\bibitem[{Wizinowich {et~al.}(2022)Wizinowich, Lu, Cetre, Chin, Correia,
  Delorme, Gers, Lilley, Lyke, Marin, {et~al.}}]{wizinowich2022keck}
Wizinowich, P., Lu, J., Cetre, S., {et~al.} 2022, in Adaptive Optics Systems
  VIII, Vol. 12185, SPIE, 193

\bibitem[{{Wizinowich} {et~al.}(2006){Wizinowich}, {Le Mignant}, {Bouchez},
  {Campbell}, {Chin}, {Contos}, {van Dam}, {Hartman}, {Johansson}, {Lafon},
  {Lewis}, {Stomski}, {Summers}, {Brown}, {Danforth}, {Max}, \&
  {Pennington}}]{Wizinowich2006}
{Wizinowich}, P.~L., {Le Mignant}, D., {Bouchez}, A.~H., {et~al.} 2006,
  \href{http://dx.doi.org/10.1086/499290}{\JournalTitle{\pasp}, 118, 297}

\end{thebibliography}
	
	\appendix
	\section{Misidentified target star}
	\label{app:A}
	
	As mentioned in the paper, the initial analysis conducted with the 2020 data was wrong since the target star had been misidentified. After submitting the paper with the wrong target star, the problem was brought to the author's attention by \cite{Gould2023}. We correct this misidentification and present the results in this paper. In this Appendix, we show the 2020 analysis conducted on the wrong target, along with new data obtained with Keck in 2023. Image reduction and the PSF fitting method is the same as what is described in Section\ref{Keck_obs}. The only difference is the target star which is being analysed. 
	
\begin{figure}[h!]
		\centering
		\includegraphics[width=12cm]{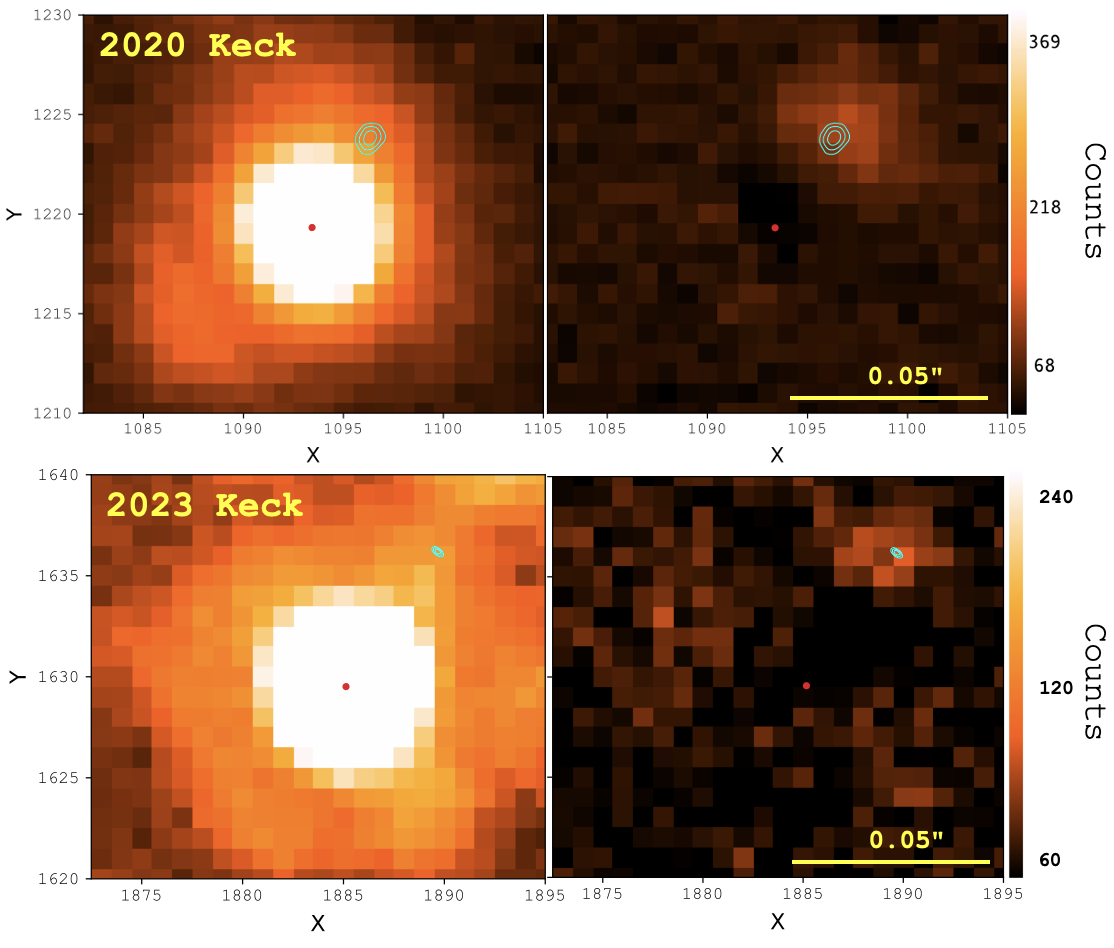}
		\caption{The 2020 and 2023 Keck narrow image data of the misidentified target star, and the 1-star fit residual image. The best fit MCMC contours of the two stars' pixel positions (X, Y) are shown in cyan (Star B) and in red (Star A). North is up and East is to the left. }
		\label{fig:res-old}
	\end{figure}

	For both the 2020 and 2023 epochs we see a visible residual in the 1-star fits (when the target star has been subtracted). We name the visible bright target `Star A' and the fainter residual star `Star B'. Star B is North West of Star A as seen in Figure \ref{fig:res-old}. 
	
	In the 2020 Keck data we find a separation between Star A and Star B of $S = 54.49 \pm 2.36$ mas, and a flux ratio of $f_{\rm B}/ f_{\rm A} = 0.065 \pm 0.007$. If Star A and Star B were in fact source and lens, then 4.12 years would have passed since the peak of the microlensing event and the relative proper motion between these two stars could be calculated to be $\mu_{\rm rel} = 13.22 \pm 0.94\ \rm mas\ yr^{-1}$.  \cite{Bond2017} predicted a relative proper motion of $\sim 9.38 \pm 0.75$ $\rm mas yr^{-1}$ from the light-curve model. Although this is a difference of $\sim 3\sigma$, light-curve models without constraints from high resolution follow-up data or robust higher order effect measurements such as parallax or xallarap, cannot always predict physical properties of the system accurately.  In addition, with the flux ratio measured and the total K magnitude of the target from the Keck 2018 wide image ($K_{\rm A+B} = 16.92 \pm 0.02$), we calculate $K_{\rm StarA,2020} = 16.98 \pm 0.05$ and $K_{\rm StarB,2020} = 19.96 \pm 0.15$. Star A is indicated in red and Star B in cyan in Figure \ref{fig:res-old}. \cite{Bond2017} predict a source K magnitude of $\sim 17.3$, which agrees with our Star A magnitude value, if we were to consider this the source star. Since the target's magnitudes agreed so well with that predicted in the light-curve model, we felt confident that we were working on the correct target star. However, this agreement was just a coinscidence, and the true source star for OB161195 was $\sim$1 magnitude brighter than predicted.
	
	In the 2023 data we measure a separation of $S = 79.71 \pm 0.49$ mas and a flux ratio of $f_{\rm B}/ f_{\rm A} = 0.132 \pm 0.004$ between the two stars. Using these two epoch's of data we can calculate the star's relative position to eachother in 2016 using the relative proper motion seen in the 3 years between 2020 and 2023. At the peak of the microlensing event, the separation should be 0 mas. Therefore, we find a relative proper motion of $\mu_{\rm rel} = 8.41 \pm 0.81 \rm mas\ yr^{-1}$, which places Star B at a separation of $\sim 16-20$ mas from Star A in 2016. Using the flux ratio found in 2023, and the calibrated total magnitude from the 2018 Keck wide data, we find $K_{\rm StarA,2023} = 17.06 \pm 0.02$ and $K_{\rm StarB,2023} = 19.25 \pm 0.03$.
	
	From the OGLE identification of OB161195 that we present in this paper, and the \cite{Gould2023} identification, we know that these stars (Star A and Star B) are unrelated nearby stars to the microlensing system OB161195.

\end{document}